\documentclass[letterpaper,onecolumn,amsmath,showpacs,preprintnumbers,eqsecnum,superscriptaddress,aps,nofootinbib,12pt,CJKutf8,ctexart]{revtex4-2}

\usepackage{mathrsfs}
\usepackage{amsfonts}
\usepackage{amsmath,amssymb,bm}
\usepackage{array}
\usepackage{verbatim}
\usepackage{epsfig}
\usepackage{graphicx}
\usepackage{hyperref}
\hypersetup{colorlinks, linkcolor = [rgb]{0, 0, 0.5}, citecolor = [rgb]{0,0.0,0.5}, urlcolor = [rgb]{0,0.0,0.5}}
\usepackage[normalem]{ulem}
\usepackage{xcolor}
\usepackage{ulem}
\usepackage{multirow}

\usepackage{CJKutf8}

\usepackage{graphicx}
\graphicspath{{./figs/}}
\usepackage{dcolumn}
\usepackage{bm}
\usepackage{slashed}
\usepackage{siunitx}
\usepackage{ulem,xpatch}
\usepackage{hyperref}
\usepackage[mathlines]{lineno}
\usepackage{comment}
\usepackage{soul}
\usepackage{xcolor}
\usepackage{xfrac}

\allowdisplaybreaks[1]

\begin{document}

\begin{CJK*}{UTF8}{gbsn}


\title{Spin Density Matrix for $\Omega^{-}$ and its Polarization Alignment in $\psi(3686)\rightarrow\Omega^{-}\bar{\Omega}^{+}$}

\newcommand*{\SDU}{Key Laboratory of Particle Physics and Particle Irradiation (MOE), Institute of Frontier and Interdisciplinary Science, Shandong University, Qingdao, Shandong 266237, China}\affiliation{\SDU}
\newcommand*{\HTU}{School of Physics, He'nan Normal University, Xinxiang, Henan 453007, China}\affiliation{\HTU}

\author{Zhe Zhang(张哲)}\email{zhangzhe@mail.sdu.edu.cn}\affiliation{\SDU}
\author{Jiao Jiao Song(宋娇娇)}\email{songjiaojiao@ihep.ac.cn}\affiliation{\HTU}

\begin{abstract}

We investigate the spin density matrix of $\Omega^{-}$ in the Cartesian coordinate system of baryon-antibaryon pairs produced in $e^{+}e^{-}$ annihilation. Using the helicity formalism of Jacob and Wick, we derive the expression for the spin-3/2 density matrices. Our analysis is based on the angular distribution of the process $e^{+}e^{-}\rightarrow\psi(3686)\rightarrow\Omega^{-}\bar{\Omega}^{+}$ in the BESIII experiment. By decomposing the polarization state of $\Omega^{-}$ particles along different coordinate axes, we examine the polarization dependence of the cross-section. Our results demonstrate that $\Omega^{-}$ particles exhibit varying degrees of tensor polarization along the $x$-, $y$-, and $z$-axes, as well as weak vector polarization and rank-3 tensor polarization along the $y$-axis. To our knowledge, this is the first study to calculate the polarization dependence of the cross-section distributions for the annihilation process $e^+e^-\rightarrow\Omega^-\bar{\Omega}^+$. Our theoretical predictions are in good agreement with the experimental measurements.\newline

\end{abstract}

\maketitle

\section{Introduction}
\label{s.intro}

Since Rutherford's discovery of the nucleus in the last century~\cite{Rutherford:1911zz}, it has been found that nucleons constitute the majority of the visible mass in the universe. As the investigation of nucleon structure involves the non-perturbative nature of strong interactions, comprehending the quarks confinement mechanism inside hadrons is still deemed as one of the most challenging tasks in contemporary physics. For strange particles, such as the $\Xi^{-}$ and $\Omega^{-}$ particles, where the light quarks inside the particles are replaced by heavier strange quarks, it follows that these heavier quarks move more slowly and exhibit simpler dynamical characteristics. Therefore, theoretical predictions for strange systems are typically more straightforward, such as calculations using the Lattice lattice QCD~\cite{Cooke:2013qqa}. The $\Omega^{-}$ baryon represents an extreme case among strange hadrons, as it consists solely of strange quarks. As a result, it plays a critical role in our comprehension of the microcosm: its discovery~\cite{Barnes:1964pd} validated the SU(3) flavor symmetry~\cite{Gell-Mann:1962yej,Gell-Mann:1964ewy} and catalyzed the development of the color charge hypothesis~\cite{Greenberg:1964pe}. Despite more than half a century having passed, our understanding of various properties of the $\Omega^-$ particle, such as spin, parity, decay parameters, and so on, remains limited.

Spin is one of the most fundamental properties of particles, and the investigation of spin structures has played a pivotal role in advancing our understanding of strong interactions and particle structures. Extensive research has been conducted on vector polarization for spin-1/2 particles. When considering the polarization of particles in the initial state, nucleons are particularly suitable for such investigations. This line of research encompasses studies on nucleon spin structure~\cite{Filippone:2001ux,Aidala:2012mv,Deur:2018roz}, distribution functions~\cite{Angeles-Martinez:2015sea,Ethier:2020way}, and other related aspects. However, when examining the polarization of particles in the final state, such as in fragmentation functions~\cite{Metz:2016swz} and global polarization studies~\cite{Liang:2004ph}, direct measurement of their polarization becomes challenging. Consequently, decay processes are often employed as a means to infer particle polarization. The $\Lambda$ hyperon, characterized by its weak decay properties, represents an ideal candidate for exploring the polarization of spin-1/2 particles in the final state. We are aware that the weak decay of spin-1/2 particles relies on decay parameters, denoted as $\alpha_\pm$, which can be measured through the $e^+e^-\rightarrow\Lambda\bar{\Lambda}$ process. In particular, the involvement of intermediate resonances such as $J/\psi$ and $\psi(2S)$ is highly suitable for precisely measuring these decay parameters, where $J/\psi$ and $\psi(2S)$ have quantum numbers $J^{PC}=1^{--}$. The cross-section for this process has been elucidated in~\cite{Faldt:2013gka,Faldt:2016qee,Faldt:2017kgy}. Through this process, the BESIII experiment has published accurate measurements of the decay parameters with high statistical precision and minimal background contributions~\cite{BESIII:2018cnd,BESIII:2021ypr}. Furthermore, the examination of disparities in decay parameters between $\Lambda$ and $\bar{\Lambda}$ provides a valuable means to quantify CP violation, which is instrumental in comprehending the matter-antimatter asymmetry observed in the universe~\cite{wuli}.

The consideration of rank-2 tensor polarization(quadrupole polarization)~\cite{Bacchetta:2000jk,Doncel:1972ez,Dubnickova:1992ii} arises when examining spin-1 particles. Recently, there has been a growing emphasis on theoretical investigations~\cite{Bacchetta:2000jk, Wei:2013csa, Chen:2020pty, Kumano:2020ijt} and experimental measurements~\cite{ALICE:2019aid, STAR:2022fan} pertaining to spin-1 particles. These endeavors have yielded invaluable insights into the structures of these particles. The $\Omega$ hyperon, being a spin-3/2 particle, exhibits vector, rank-2 tensor(quadrupole), and rank-3 tensor(octupole) polarization states~\cite{Song:1967,Kim:1976dn,Doncel:1972ez,Dubnickova:1992ii}. Due to its weak decay characteristics, akin to the $\Lambda$ particle, these polarizations can be probed via decay chains, offering a novel perspective on particle spin structures. Recently, the BESIII experiment published measurement results on the polarization states of $\Omega$ particles generated in the $e^+e^-\rightarrow\psi(3868)\rightarrow\Omega^-\bar{\Omega}^+$ process~\cite{BESIII:2020lkm}, employing the helicity formalism as developed in~\cite{Jacob:1959at,Perotti:2018wxm}. These results revealed intriguing polarization-dependent phenomena in the reaction cross-section. To comprehend these experimental findings, further theoretical progress in spin-3/2 particle studies is imperative. Several theoretical articles addressing spin-3/2 particles have already been published~\cite{Song:1967,Kim:1976dn,Yang:2020ezt,Zhao:2022lbw,Doncel:1972ez,Perotti:2018cpb,Perotti:2018wxm}. Nonetheless, it is noteworthy that various conventions exist, which may introduce confusion. For instance, the definition of polarization components in the helicity formulation decomposition~\cite{Perotti:2018wxm} may not coincide with those defined by the spin density matrix~\cite{Zhao:2022lbw}. Although their essential nature remains unchanged, these discrepancies can lead to confusion and unwarranted complexities. Additionally, the calculation of cross-sections for spin-3/2 particle production processes remains an area that necessitates further development.

In this paper, we commence by providing an overview of the helicity formalism utilized in the $e^+e^-\rightarrow\psi(3868)\rightarrow\Omega^-\bar{\Omega}^+$ process and the decomposition of polarization components for spin-3/2 particles. We redefine the polarization basis matrices within the helicity formalism, establishing a direct correspondence between the decomposition of polarization components in this formalism and the definition of polarization components in the spin density matrix. This connection enables us to establish a direct relationship between experimentally measured parameters and specific interpretations of polarization. For spin-3/2, there exist four polarization states along any given direction, with these polarization components representing intricate combinations of probabilities for different polarization states along various directions. The complexity involved poses challenges to our intuitive understanding of the polarization of the $\Omega^-$ particle. To address this, we project the polarization states of the $\Omega^-$ particle onto the $x$-, $y$-, and $z$-axes in a Cartesian coordinate system, thereby providing a clear visualization of the polarization states along each axis. Our results reveal that $\Omega^-$ particles exhibit varying degrees of tensor polarization along the $x$-, $y$-, and $z$-axes, as well as weak vector polarization and rank-3 tensor polarization along the $y$-axis.

Furthermore, in order to account for the polarization dependence of the cross-section on azimuthal angle $\theta_{\Omega}$, we conducted the first calculation of the cross-section for the process $e^+e^-\rightarrow\Omega^-\bar{\Omega}^+$. By neglecting the form factors, we are focusing solely on the spin-3/2 field equation as a means to describe the polarization state of the $\Omega^-$ particle. This investigation allows us to assess the extent to which the field equation alone can capture the intricate polarization properties of the particle, shedding light on the role and limitations of the field equation in characterizing its polarization behavior. Overall, our theoretical calculations demonstrate good agreement with experimental results.

The remaining sections of the paper are organized as follows. In Sec.~\ref{P:production}, we presents the general helicity framework, which has been adjusted to fit commonly employed experimental analysis. In Sec.~\ref{S:sdm}, we construct the spin density matrix of baryons that we are interested in, i.e. spin-1/2 and spin-3/2 baryons. In Sec.~\ref{s.decomposition}, we elucidate the polarization states of the $\Omega^-$ particle by projecting them onto the coordinate axes in a Cartesian coordinate system.  In Sec.~\ref{s.cross}, we calculate the cross-section for process $e^+e^-\rightarrow\Omega^-\bar{\Omega}^+$ and compare the results with experimental measurements. A brief summary will be given in Sec.~\ref{S:summary}.

\section{Experimental production process}\label{P:production}

\begin{figure}
\centering
   \includegraphics[width=0.50\textwidth]{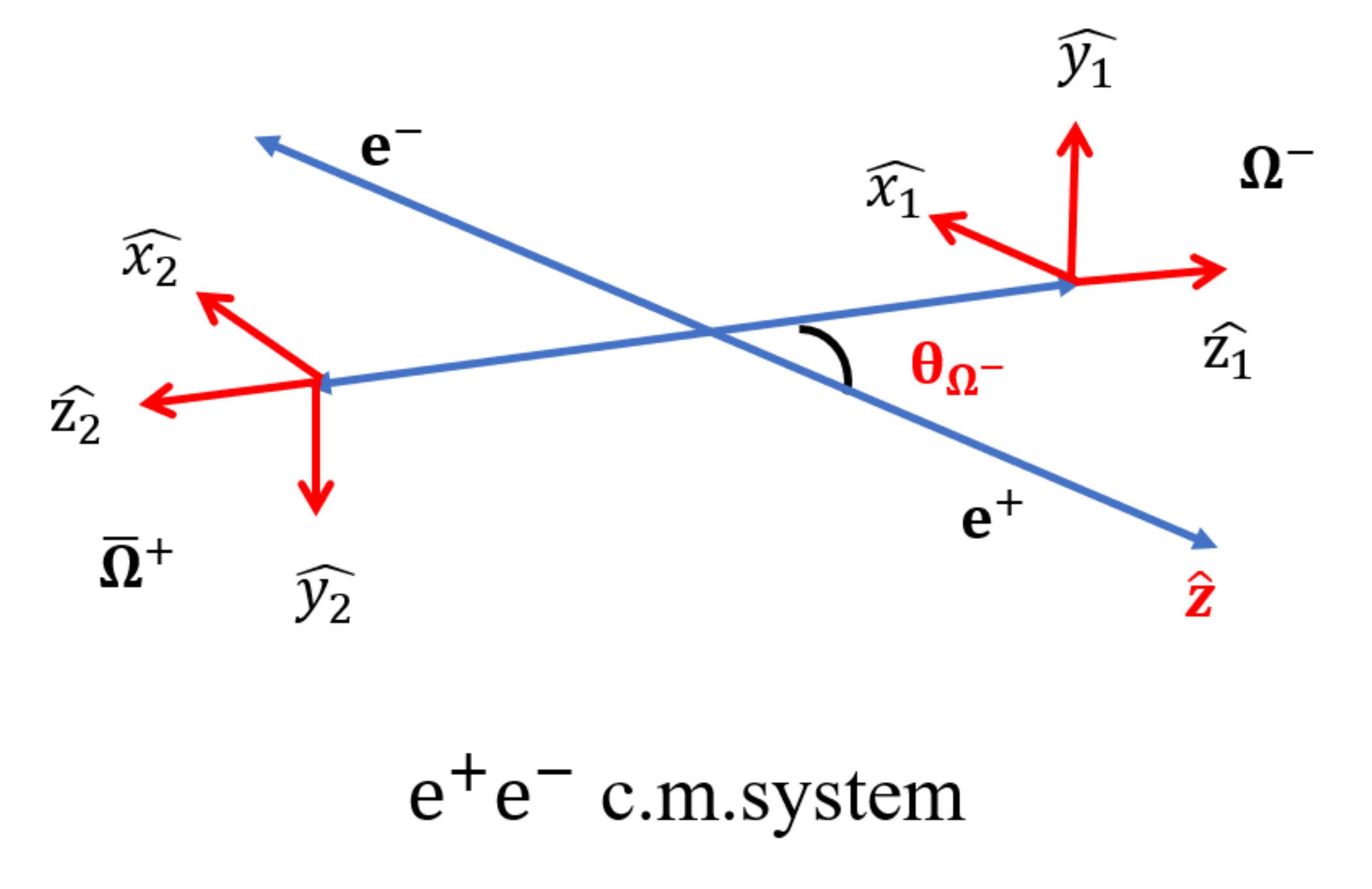}\\
\caption{In the helicity formalism, the helicity angle $\theta_{\Omega^{-}}$ is defined as the polar angle of the $\Omega^{-}$ particle in the center-of-mass (c.m.) system of the $e^+e^-$ collision. The $\widehat{z}$-axis in the $e^+e^-$ c.m. system corresponds to the direction of the incoming positron.} \label{helicity_angle}
\end{figure}

We describe the production amplitudes of $e^{+}e^{-}$ annihilation into $\Omega^{-}$ $\bar{\Omega}^{+}$ hyperons using the helicity formalism introduced by Jacob and Wick in their seminal work~\cite{Jacob:1959at}. In this formalism, the helicity angle $\theta_{\Omega^{-}}$ corresponds to the polar angle of the $\Omega^{-}$ particle in the center-of-mass (c.m.) frame of the $e^{+}e^{-}$ collision and serves as the variable of interest. The helicity system for the $\Omega^{-}$ $\bar{\Omega}^{+}$ production process is defined and illustrated in Fig.~\ref{helicity_angle}.

In the helicity framework, the production process of $e^+e^-\rightarrow\gamma^*\rightarrow\psi(3686)\rightarrow\Omega^-\bar{\Omega}^+$ can be described as~\cite{Perotti:2018wxm}:
\begin{equation}\label{helicity_density}
\begin{aligned}
\rho^{\lambda_{1},\lambda_{2};\lambda_{1}^{'},\lambda_{2}^{'}}\propto
A_{\lambda_{1},\lambda_{2}}A^{*}_{\lambda_{1}^{'},\lambda_{2}^{'}}
\rho^{\lambda_{1}-\lambda_{2},\lambda_{1}^{'}-\lambda_{2}^{'}}_{1}(\theta_{\Omega^-}),\\
\rho^{i,j}_{1}:=\sum_{\kappa=\pm1}D^{1*}_{\kappa,i}(0,\theta_{\Omega^-},0)D^{1}_{\kappa,j}(0,\theta_{\Omega^-},0),
\end{aligned}
\end{equation}
where $D$ represents the Wigner transformation matrix, $\theta_{\Omega^{-}}$ is the helicity angle of the $\Omega^{-}$ particle, $A_{\lambda_1,\lambda_2}$ ($A_{\lambda_1',\lambda_2'}$) denotes the transition amplitude with helicities $\lambda_1$ ($\lambda_1'$) and $\lambda_2$ ($\lambda_2'$) for the $\Omega^{-}$ and $\bar{\Omega}^{+}$ particles, respectively, $\kappa$ represents the helicity difference between the initial $e^{+}$ and $e^{-}$ states, which can only take values of $\pm 1$. Since the $e^{+}e^{-}$ beam is non-polarized, it is necessary to sum over the helicity difference $\kappa$. If only the $\Omega^{-}$ particle is reconstructed, without considering the decays of the recoil side, the density function for a single $\Omega^{-}$ particle can be obtained by requiring $\lambda_2 = \lambda_{2}^{\prime}$ and summing over the helicity states:
\begin{equation}\label{single_helicity_density}
\begin{aligned}
\rho_{\Omega^{-}}\propto \sum_{\lambda_{2}}A_{\lambda_{1},\lambda_{2}}A^{*}_{\lambda_{1}^{'},\lambda_{2}}\rho_{1}^{\lambda_{1}-\lambda_{2},\lambda_{1}^{'}-\lambda_{2}}(\theta_{\Omega^{-}}).
\end{aligned}
\end{equation}

Due to the conservation of parity and charge conjugate invariance, only four transition amplitudes are obtained:


\begin{equation}
\begin{aligned}
H_1 :=& A_{\frac{1}{2},\frac{1}{2}}=A_{-\frac{1}{2},-\frac{1}{2}},\\
H_2 :=& A_{\frac{1}{2},-\frac{1}{2}}=A_{-\frac{1}{2},\frac{1}{2}},\\
H_3 :=& A_{\frac{3}{2},\frac{1}{2}}=A_{-\frac{3}{2},-\frac{1}{2}}\\=&A_{\frac{1}{2},\frac{3}{2}}=A_{-\frac{1}{2},-\frac{3}{2}},\\
H_4 :=& A_{\frac{3}{2},\frac{3}{2}}=A_{-\frac{3}{2},-\frac{3}{2}}.\\
\end{aligned}
\end{equation}
And the transition amplitude matrix is given by:
\begin{equation}\label{amplitude_matrix}
 \left(
 \begin{array}{cccc}
   H_4 & H_3 & 0 &0\\
   H_3 & H_1 & H_2 &0\\
   0   & H_2 & H_1 & H_3\\
   0   & 0   & H_3 & H_4\\
  \end{array}
  \right).
  \end{equation}


In the experimental data analysis, if one only cares about the azimuthal dependence of the reaction cross-section and polarization, one can consider the ratios of the amplitudes without considering their absolute magnitudes. These ratios are defined as follows:
$H_1/H_2=h_1e^{i\phi_1}$,
$H_3/H_2=h_3e^{i\phi_3}$,
$H_4/H_2=h_4e^{i\phi_4}$,
where $h_i$ and $\phi_i$ (i=1,3,4) are real variables obtained through experimental fitting.
Based on the aforementioned theoretical framework, BESIII has conducted an analysis of the angular distribution of the single-tag process $\psi(3686)\to \Omega^{-}\bar{\Omega}^{+}$ ($\Omega^-\to K^-\Lambda$, $\Lambda\to p\pi^-$) using a data set of $448\times 10^6$ $\psi(3686)$ decays collected with the BESIII detector at the BEPCII electron-positron collider~\cite{BESIII:2020lkm}. The analysis results are presented in Table~\ref{simultaneous_result}.
\begin{table}[htbp]
 \centering
\caption{Two sets of fit values of the helicity parameters of $\psi(3686)\to \Omega^{-}\bar{\Omega}^{+}$ decays in BESIII measurement~\cite{BESIII:2020lkm}. The first uncertainties are statistical, and the second ones systematic.}\label{simultaneous_result}
   \begin{tabular}{c|c|c}
   \hline\hline
   parameter & solution I & solution II  \\\hline
   $h_1$           & 0.30$\pm$0.11$\pm$0.04   & 0.31$\pm$0.10$\pm$0.04\\
   $\phi_1$        & 0.69$\pm$0.41$\pm$0.13   & 2.38$\pm$0.37$\pm$0.13\\
   $h_3$           & 0.26$\pm$0.05$\pm$0.02   & 0.27$\pm$0.05$\pm$0.01\\
   $\phi_3$        & 2.60$\pm$0.16$\pm$0.08   & 2.57$\pm$0.16$\pm$0.04\\
   $h_4$           & 0.51$\pm$0.03$\pm$0.01   & 0.51$\pm$0.03$\pm$0.01\\
   $\phi_4$        & 0.34$\pm$0.80$\pm$0.31   & 1.37$\pm$0.68$\pm$0.16\\
   \hline \hline
   \end{tabular}
\end{table}

\section{Baryon spin density matrices}\label{S:sdm}

The description of particles with spin can be achieved using a spin density matrix $\rho$ in the rest frame of the particle, as discussed in the literature~\cite{Leader:2011vwq}. The spin density matrix enables the characterization of the spin state of a system by introducing irreducible spin components up to rank $2J$. For example, the most general spin density matrix for a spin-1/2 particle is given by the following form, as described in~\cite{Leader:2011vwq}:
\begin{align}
  \rho_{1/2} = \frac{1}{2}(\bm{1}+ S^i\mathbf{\sigma}^i), \label{f:rhocartesin}
\end{align}
where $\mathbf{\sigma}^i$ represents the set of Pauli matrices. The (rank-one) spin vector $S^i$ corresponds to the transverse ($S_{x}$, $S_{y}$) and longitudinal ($S_{L}$) vector polarizations of the particle.

To parametrize the density matrix of a spin-3/2 particle, we can utilize a Cartesian basis comprising 4 × 4 matrices. This basis includes the identity matrix and three spin matrices denoted as $\Sigma^i$, which serve as a generalization of the Pauli matrices to the four-dimensional case:
 \begin{equation}\label{f:spinvector}
\begin{aligned}
  &\Sigma^x=\frac{1}{2}\left(
        \begin{array}{cccc}
          0 & \sqrt{3} & 0 & 0 \\
          \sqrt{3} & 0 & 2 & 0 \\
          0 & 2 & 0 & \sqrt{3} \\
          0 & 0 & \sqrt{3} & 0 \\
        \end{array}
      \right),
 \quad  \Sigma^y=\frac{i}{2}\left(
        \begin{array}{cccc}
          0 & -\sqrt{3} & 0 & 0 \\
          \sqrt{3} & 0 & -2 & 0 \\
          0 & 2 & 0 & -\sqrt{3} \\
          0 & 0 & \sqrt{3} & 0 \\
        \end{array}
      \right),
 \quad \Sigma^z=\frac{1}{2}\left(
                   \begin{array}{cccc}
                     3 & 0 & 0 & 0 \\
                     0 & 1 & 0 & 0 \\
                     0 & 0 & -1 & 0 \\
                     0 & 0 & 0 & -3 \\
                   \end{array}
                 \right).
\end{aligned}
\end{equation}
 In addition to these, there are five extra matrices of rank-2 denoted as $\Sigma^{ij}$ and seven extra matrices of rank-3 denoted as $\Sigma^{ijk}$~\cite{Zhao:2022lbw, Song:1967}. These matrices provide a comprehensive representation for the density matrix in the $s_z$ basis. These last two can be constructed from the spin matrices of $(\Sigma^{i})s$. The (rank-2) spin tensor basis ($\Sigma^{ij}$)s are given by~:
\begin{equation}
\begin{aligned}
  &\Sigma^{ij}=\frac{1}{2}(\Sigma^i\Sigma^j+\Sigma^j\Sigma^i)-\frac{5}{4} \delta^{ij}\bm{1}. \label{f:sigmaij}
\end{aligned}
\end{equation}
And five independent spin tensor basis $\Sigma^{zx}$, $\Sigma^{zy}$, $\Sigma^{zz}$, $\Sigma^{yx}$ and ($\Sigma^{xx}-\Sigma^{yy}$) are selected. Similarly, we construct the (rank-3) spin tensor basis ($\Sigma^{ijk}$)s as:
\begin{equation}
\begin{aligned}
  \Sigma^{ijk}=\frac{1}{6}\Sigma^{\{i}\Sigma^j\Sigma^{k\}}-\frac{41}{60}(\delta^{ij}\Sigma^k+\delta^{jk}\Sigma^i
  +\delta^{ki}\Sigma^j). \label{f:sigmaijk}
\end{aligned}
\end{equation}
where the symbol $\{\cdots\}$ represents the symmetrization of the indices, indicating the sum over all permutations. And seven independent spin tensor basis $\Sigma^{xzz}$, $\Sigma^{yzz}$, $\Sigma^{zzz}$, $\Sigma^{xyz}$, ($\Sigma^{xxz}-\Sigma^{yyz}$), ($\Sigma^{xxx}-3\Sigma^{xyy}$) and (3$\Sigma^{xxy}-\Sigma^{yyy}$) are selected.

With these preliminaries, the spin density matrix can be expressed as follows~\cite{Zhao:2022lbw}:
\begin{equation}
\begin{aligned}
  \rho_{3/2}=\frac{1}{4}(\bm{1} + \frac{4}{5}S^i\Sigma^i + \frac{2}{3}T^{ij}\Sigma^{ij} + \frac{8}{9}R^{ijk}\Sigma^{ijk}), \label{f:rhomatrix}
\end{aligned}
\end{equation}
where we introduce the symmetric traceless (rank-2) spin tensor $T^{ij}$ and the (rank-3) spin tensor $R^{ijk}$. The spin density matrix consists of fifteen spin components, which are given by
 \begin{equation}\label{spin_components}
\begin{aligned}
 S^{i}&: S_{L}, S^{x}_{T}, S^{y}_{T},\\
 T^{ij}&: S_{LL}, S^{x}_{LT}, S^{y}_{LT}, S_{TT}^{xx}, S_{TT}^{xy},\\
 R^{ijk}&: S_{LLL}, S^{x}_{LLT}, S^{y}_{LLT},S_{LTT}^{xx}, S_{LTT}^{xy},  S_{TTT}^{xxx}, S_{TTT}^{yxx}.\\
\end{aligned}
\end{equation}
These spin components can be described as the probabilities of finding specific eigenstates from the spin density matrix $\rho_{3/2}$, as detailed in Appendix~\ref{A:Probabilistic}. The domains of the spin components can be determined through probabilistic interpretations:
\begin{equation}\label{spin_components_value}
\begin{aligned}
 &S_{L}, S^{x}_{T}, S^{y}_{T}\in[-\frac{3}{2},\frac{3}{2}],\\
 &S_{LL}\in[-1,1],\quad S^{x}_{LT}, S^{y}_{LT}, S_{TT}^{xy}, S_{TT}^{xx}\in[-\sqrt{3},\sqrt{3}],\\
 &S_{LLL}\in[-\frac{9}{10},\frac{9}{10}],\quad S^{x}_{LLT}, S^{y}_{LLT}\in[-\frac{3+\sqrt{21}}{10},\frac{3+\sqrt{21}}{10}],\\
 &S_{LTT}^{xx},S_{LTT}^{xy}\in[-\sqrt{3},\sqrt{3}],\quad S_{TTT}^{xxx}, S_{TTT}^{yxx}\in[-3,3].\\
\end{aligned}
\end{equation}

To simplify the notation, we adopt a shorthand notation where a single index is used to represent the spin matrices:
\begin{equation}
\begin{aligned}
 \Sigma_0 &=\frac{1}{4}\bm{1}, \\
\Sigma_1 &=\frac{1}{5} \Sigma^z,\quad \Sigma_2=\frac{1}{5} \Sigma^x,\quad \Sigma_3=\frac{1}{5} \Sigma^y ,\\
\Sigma_4 &=\frac{1}{4} \Sigma^{z z},\quad \Sigma_5=\frac{1}{6} \Sigma^{x z},\quad \Sigma_6=\frac{1}{6} \Sigma^{y z},\quad \Sigma_7=\frac{1}{12}\left(\Sigma^{x x}-\Sigma^{y y}\right),\quad \Sigma_8=\frac{1}{6} \Sigma^{x y}, \\
\Sigma_9 &=\frac{5}{9} \Sigma^{z z z},\quad \Sigma_{10}=\frac{5}{6} \Sigma^{x z z},\quad \Sigma_{11}=\frac{5}{6} \Sigma^{y z z},\quad \Sigma_{12}=\frac{1}{6}\left(\Sigma^{x x z}-\Sigma^{y y z}\right), \\
\Sigma_{13} &=\frac{1}{3} \Sigma^{x y z}, \quad\Sigma_{14}=\frac{1}{18}\left(\Sigma^{x x x}-3 \Sigma^{x y y}\right), \quad\Sigma_{15}=\frac{1}{18}\left(3 \Sigma^{x x y}-\Sigma^{y y y}\right),
\end{aligned}
\end{equation}
 where $\bm{1}$ is the 4 $\times$ 4 identity matrix. The explicit expression for these basis matrices is given in Appendix~\ref{B:Matrices}. Then the corresponding spin density matrix of a spin-3/2 particle can be expressed as follows using the shorthand notation:
\begin{equation}\label{Rho_3/2}
\begin{aligned}
  \rho_{3/2}=\Sigma_{\mu=0}^{15}S_{\mu}\Sigma_{\mu},
\end{aligned}
\end{equation}
where $S_{0}$ represents the cross-section term, and $S_{\mu}$ are 15 real numbers that correspond to the fifteen spin components. By using the chosen matrix basis, we have established a direct mapping between the decomposition of spin components in the helicity formalism and the decomposition of spin components in terms of the density matrix~\cite{Zhao:2022lbw}. The relationship between the polarization components $S_{L}$, $S^{x}_{T}$ $\cdots$ and the expansion coefficients $S_{\mu}$ is presented in Table~\ref{correspond_coefficient}.
\begin{table}[htbp]
 \centering
\caption{The correspondence of the expansion coefficient $S_{\mu}$ and the spin components.}\label{correspond_coefficient}
   \begin{tabular}{c|c|c|c|c|c|c|c|c|c|c|c|c|c|c|c}
   \hline\hline
    $1$&$S_{L}$ & $S^{x}_{T}$ &$S^{y}_{T}$ &$S_{LL}$ &$S^{x}_{LT}$ &$S^{y}_{LT}$ &$S^{xx}_{TT}$ &$S^{xy}_{TT}$ &$S_{LLL}$  &$S^{x}_{LLT}$   &$S^{y}_{LLT}$  &$S^{xx}_{LTT}$ &$S^{xy}_{LTT}$ &$S^{xxx}_{TTT}$ &$S^{yxx}_{TTT}$   \\\hline
     $S_{0}$ &$S_{1}$ & $S_{2}$ &$S_{3}$ &$S_{4}$ &$S_{5}$ &$S_{6}$ &$S_{7}$ &$S_{8}$ &$S_{9}$  &$S_{10}$ &$S_{11}$ &$S_{12}$ &$S_{13}$ &$S_{14}$ &$S_{15}$   \\
   \hline\hline
   \end{tabular}
\end{table}

The density matrix of $\Omega^{-}$ can also be expressed in terms of the helicity amplitudes as given in Eq.~\eqref{single_helicity_density}. Due to the parity conserving nature of the process, it is found that only seven parameters are non-zero, namely $S_{0}$, $S_{3}$, $S_{4}$, $S_{5}$, $S_{7}$, $S_{11}$, and $S_{13}$:
\begin{equation}\label{coefficient_S}
\begin{aligned}
  S_0 &=2 \sin ^2 \theta_{\Omega^{-}}\left(\left|H_1\right|^2+\left|H_4\right|^2\right)+\left(1+\cos ^2 \theta_{\Omega^{-}}\right)\left(\left|H_2\right|^2+2\left|H_3\right|^2\right) ,\\
S_3 &=\frac{1}{\sqrt{2}} \sin 2 \theta_{\Omega^{-}}\left(2 \operatorname{Im}\left[H_2 H_1^*\right]+\sqrt{3} \operatorname{Im}\left[H_3\left(H_1^*+H_4^*\right)\right]\right) ,\\
S_4 &=2 \sin ^2 \theta_{\Omega^{-}}\left(\left|H_4\right|^2-\left|H_1\right|^2\right)-\left(1+\cos ^2 \theta_{\Omega^{-}}\right)\left|H_2\right|^2 ,\\
S_5 &=\sqrt{6} \sin 2 \theta_{\Omega^{-}} \operatorname{Re}\left[\left(H_4-H_1\right) H_3^*\right], \\
S_7 &=2 \sqrt{3} \sin ^2 \theta_{\Omega^{-}} \operatorname{Re}\left[H_2 H_3^*\right] ,\\
S_{11} &=\frac{\sqrt{2}}{5} \sin \left(2 \theta_{\Omega^{-}}\right)\left(3 \operatorname{Im}\left[H_1 H_2^*\right]+\sqrt{3} \operatorname{Im}\left[H_3\left(H_1^*+H_4^*\right)\right]\right), \\
S_{13} &=2 \sqrt{3} \sin ^2 \theta_{\Omega^{-}} \operatorname{Im}\left[H_2 H_3^*\right].
\end{aligned}
\end{equation}
These parameters characterize the spin properties of the $\Omega^{-}$ particle in the considered process.

\begin{figure}
\centering
   \includegraphics[width=0.4\textwidth]{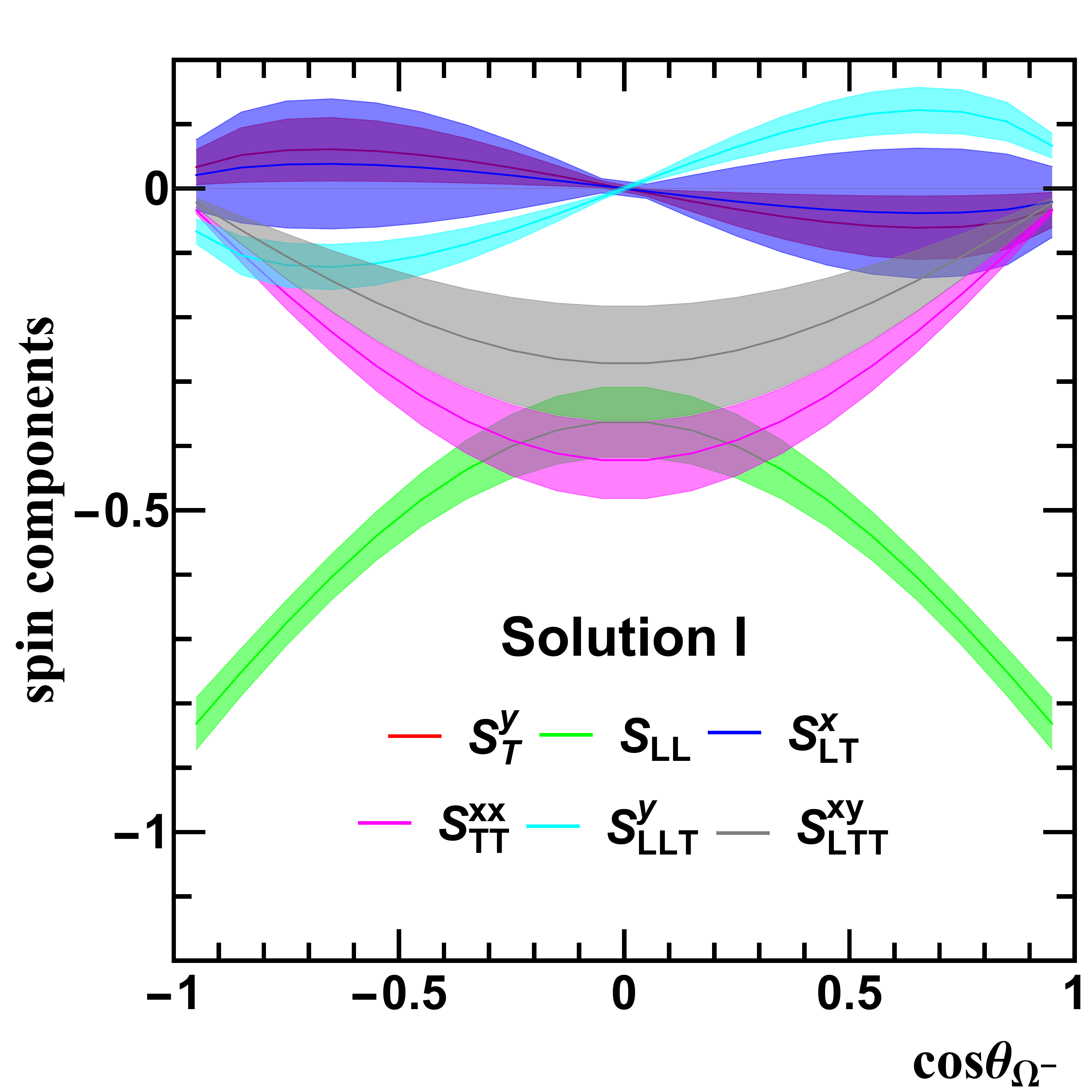}\includegraphics[width=0.4\textwidth]{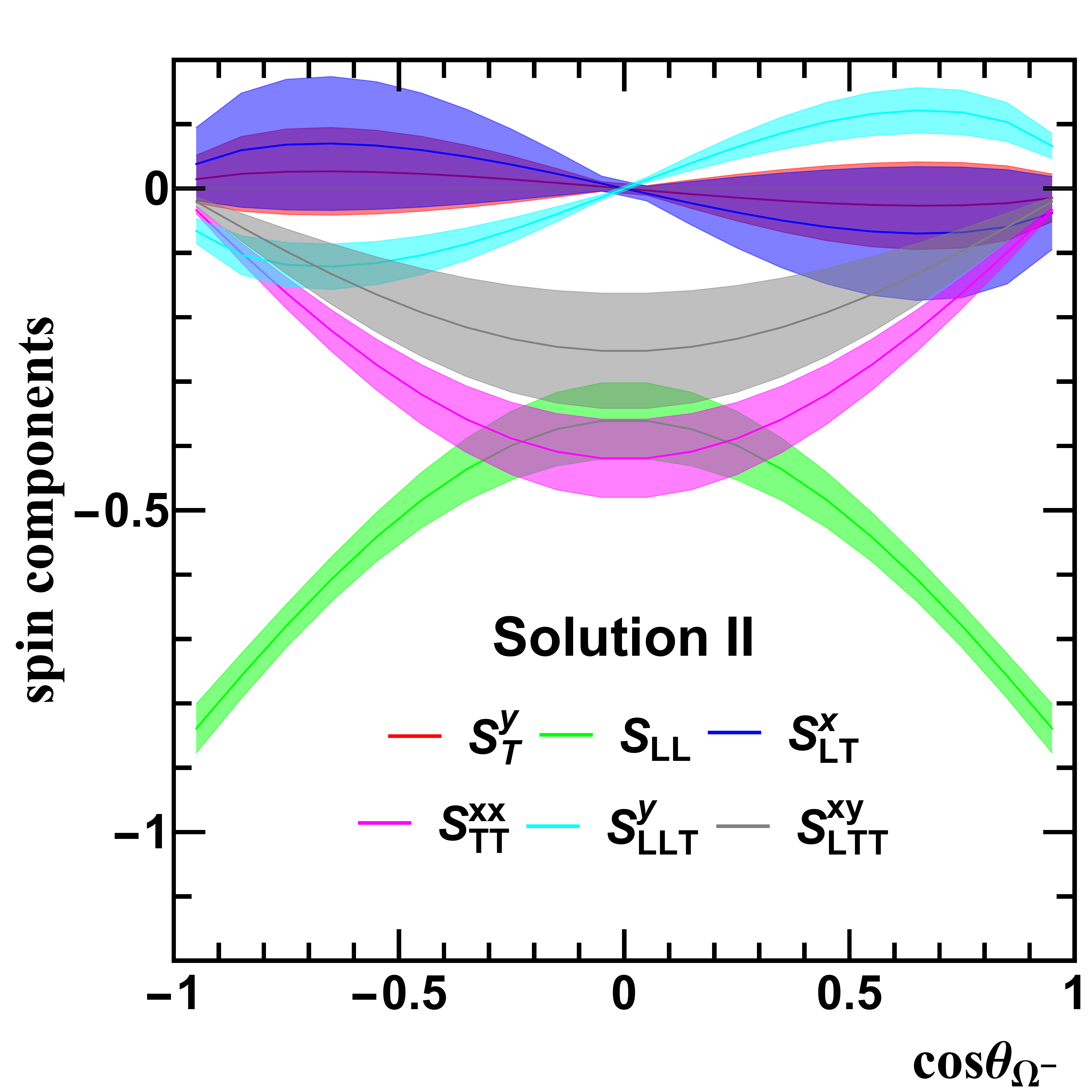}
\caption{The polarization dependence of the reaction cross-section exhibits variations with the azimuthal angle $\cos\theta_{\Omega^{-}}$. The solid lines correspond to the central values, while the shaded areas indicate the uncertainty range of $\pm$ one standard deviation.}
\label{polarization_Omega}
\end{figure}

With the measured helicity amplitudes provided in Table~\ref{simultaneous_result}, we have investigated the polarization dependence of the reaction cross-section as a function of the azimuthal angle $\cos\theta_{\Omega^{-}}$, as shown in Fig.~\ref{polarization_Omega}. The uncertainties, which include both statistical and systematic contributions, have been estimated using the experimentally determined covariance matrix of the fitted $h_i$ and $\phi_i$ parameters~\cite{BESIII:2020lkm}. In the $e^{+}e^{-}\to \psi(3686)\to \Omega^{-}\bar{\Omega}^{+}$ process, the $\Omega^{-}$ particles exhibit not only transverse vector polarization ($S_3$), but also contributions from rank-2 tensor polarization ($S_4$, $S_5$, $S_7$) and rank-3 tensor polarization ($S_{11}$, $S_{13}$). All of these spin components have been computed using the helicity amplitudes listed in Table~\ref{simultaneous_result}. Indeed, it is worth noting that both sets of solutions yield the same overall effect. For the sake of clarity and simplicity, we have chosen to present only one set of solutions in our analysis and figures.

\section{Polarization probabilistic interpretation} \label{S:probability}
In Section \ref{S:sdm}, we provide the Cartesian form of the spin density matrix for $\Omega^{-}$, which is parameterized by vector polarizations and tensor polarizations represented by the coefficients $S_{\mu}$. Each component of the spin state corresponds to a combination of probabilities, indicating the likelihood of finding the system in a specific spin state defined in the particle's rest frame. In this section, we focus on analyzing the polarization dependence of the cross-section along the coordinate axes and provide a physical interpretation for the observed behaviors.

\subsection{Polarization decomposition along coordinate axes} \label{s.decomposition}
The spin operator can be defined in any direction, and for our analysis, we choose to express it in terms of the eigenstates of the spin vector operator in a specific direction. This allows us to write the spin vector operator in terms of polar and azimuthal angles~\cite{Bacchetta:2000jk}.
\begin{align}
  \Sigma^i\hat n^i=\Sigma^x \sin\theta\cos\phi +\Sigma^y \sin\theta\sin\phi +\Sigma^z \cos\theta, \label{f:spinoperator}
\end{align}

As the common convention, we denote the eigenstate with eigenvalue $m$ along the direction $(\theta,\phi) $ as $|m_{(\theta,\phi)}\rangle $. The eigenstates of $\Sigma^z$, $\Sigma^x$ and $\Sigma^y$ are given by

\begin{equation}\label{direction_Axis}
\begin{aligned}
&\left|\frac{3}{2}\right\rangle_z=\left(\begin{array}{c}
1 \\
0 \\
0 \\
0
\end{array}\right), \,\left|\frac{1}{2}\right\rangle_z=\left(\begin{array}{l}
0 \\
1 \\
0 \\
0
\end{array}\right), \,\left|-\frac{1}{2}\right\rangle_z=\left(\begin{array}{l}
0 \\
0 \\
1 \\
0
\end{array}\right), \,\left|-\frac{3}{2}\right\rangle_z=\left(\begin{array}{c}
0 \\
0 \\
0 \\
1
\end{array}\right), \\
&\left|\frac{3}{2}\right\rangle_x=\frac{\sqrt{2}}{4}\left(\begin{array}{c}
1 \\
\sqrt{3} \\
\sqrt{3} \\
1
\end{array}\right), \,\left|\frac{1}{2}\right\rangle_x=\frac{\sqrt{2}}{4}\left(\begin{array}{c}
\sqrt{3} \\
1 \\
-1 \\
-\sqrt{3}
\end{array}\right), \,\left|-\frac{1}{2}\right\rangle_x=\frac{\sqrt{2}}{4}\left(\begin{array}{c}
\sqrt{3} \\
-1 \\
-1 \\
\sqrt{3}
\end{array}\right), \,\left|-\frac{3}{2}\right\rangle_x=\frac{\sqrt{2}}{4}\left(\begin{array}{c}
1 \\
-\sqrt{3} \\
\sqrt{3} \\
-1
\end{array}\right), \\
&\left|\frac{3}{2}\right\rangle_y=\frac{\sqrt{2}}{4}\left(\begin{array}{c}
1 \\
i \sqrt{3} \\
-\sqrt{3} \\
-i
\end{array}\right), \,\left|\frac{1}{2}\right\rangle_y=\frac{\sqrt{2}}{4}\left(\begin{array}{c}
\sqrt{3} \\
i \\
1 \\
i \sqrt{3}
\end{array}\right), \,\left|-\frac{1}{2}\right\rangle_y=\frac{\sqrt{2}}{4}\left(\begin{array}{c}
\sqrt{3} \\
-i \\
1 \\
-i \sqrt{3}
\end{array}\right), \,\left|-\frac{3}{2}\right\rangle_y=\frac{\sqrt{2}}{4}\left(\begin{array}{c}
1 \\
-i \sqrt{3} \\
-\sqrt{3} \\
i
\end{array}\right),
\end{aligned}
\end{equation}
where $\Sigma^{i}\left|m\right\rangle_i = m \left|m\right\rangle_i$, such as $\Sigma^{z}\left|\frac{3}{2}\right\rangle_z = \frac{3}{2} \left|\frac{3}{2}\right\rangle_z$. We have used a more intuitive notation ($x$, $y$ and $z$) to denote the direction angles, the relation between the two notations is the following
\begin{equation}
|m\rangle_z=\left|m_{(0,0)}\right\rangle, \quad|m\rangle_x=\left|m_{\left(\frac{\pi}{2}, 0\right)}\right\rangle, \quad|m\rangle_y=\left|m_{\left(\frac{\pi}{2}, \frac{\pi}{2}\right)}\right\rangle.
\end{equation}
The probability of observing one of these eigenstates is defined as follows:
\begin{equation}\label{d_probability}
P_i(m)\equiv {_i}\langle m|\rho_{3/2} |m\rangle_{i}.
\end{equation}
Inserting the Eqs.~\eqref{Rho_3/2} and~\eqref{direction_Axis} into~\eqref{d_probability}, one can readily obtain:
\begin{equation}\label{xyz_probability}
\begin{aligned}
P_z\left(\frac{3}{2}\right) &=\frac{1}{4}+\frac{3}{10} S_L+\frac{1}{4} S_{L L}+\frac{1}{6} S_{L L L}=\frac{1}{4}+\frac{S_4}{4},\\
P_z\left(\frac{1}{2}\right) &=\frac{1}{4}+\frac{1}{10} S_L-\frac{1}{4} S_{L L}-\frac{1}{2} S_{L L L}=\frac{1}{4}-\frac{S_4}{4},\\
P_z\left(-\frac{1}{2}\right) &=\frac{1}{4}-\frac{1}{10} S_L-\frac{1}{4} S_{L L}+\frac{1}{2} S_{L L L}=\frac{1}{4}-\frac{S_4}{4},\\
P_z\left(-\frac{3}{2}\right) &=\frac{1}{4}-\frac{3}{10} S_L+\frac{1}{4} S_{L L}-\frac{1}{6} S_{L L L}=\frac{1}{4}+\frac{S_4}{4},\\
P_x\left(\frac{3}{2}\right) &=\frac{1}{4}+\frac{3}{10} S_T^x-\frac{1}{8} S_{L L}+\frac{1}{8} S_{T T}^{x x}-\frac{1}{8} S_{L L T}^x+\frac{1}{24} S_{T T T}^{x x x}=\frac{1}{4}-\frac{S_4}{8} +\frac{S_7}{8} ,\\
P_x\left(\frac{1}{2}\right) &=\frac{1}{4}+\frac{1}{10} S_T^x+\frac{1}{8} S_{L L}-\frac{1}{8} S_{T T}^{x x}+\frac{3}{8} S_{L L T}^x-\frac{1}{8} S_{T T T}^{x x x}=\frac{1}{4}+\frac{S_4}{8} -\frac{S_7}{8} ,\\
P_x\left(-\frac{1}{2}\right) &=\frac{1}{4}-\frac{1}{10} S_T^x+\frac{1}{8} S_{L L}-\frac{1}{8} S_{T T}^{x x}-\frac{3}{8} S_{L L T}^x+\frac{1}{8} S_{T T T}^{x x x}=\frac{1}{4}+\frac{S_4}{8} -\frac{S_7}{8} ,\\
P_x\left(-\frac{3}{2}\right) &=\frac{1}{4}-\frac{3}{10} S_T^x-\frac{1}{8} S_{L L}+\frac{1}{8} S_{T T}^{x x}+\frac{1}{8} S_{L L T}^x-\frac{1}{24} S_{T T T}^{x x x}=\frac{1}{4}-\frac{S_4}{8} +\frac{S_7}{8} ,\\
P_y\left(\frac{3}{2}\right) &=\frac{1}{4}+\frac{3}{10} S_T^y-\frac{1}{8} S_{L L}-\frac{1}{8} S_{T T}^{x x}-\frac{1}{8} S_{L L T}^y-\frac{1}{24} S_{T T T}^{y x x}=\frac{1}{4}+\frac{3 S_3}{10} -\frac{S_4}{8} -\frac{S_7}{8} -\frac{S_{11}}{8} ,\\
P_y\left(\frac{1}{2}\right) &=\frac{1}{4}+\frac{1}{10} S_T^y+\frac{1}{8} S_{L L}+\frac{1}{8} S_{T T}^{x x}+\frac{3}{8} S_{L L T}^y+\frac{1}{8} S_{T T T}^{y x x}=\frac{1}{4}+\frac{S_3}{10} +\frac{S_4}{8} +\frac{S_7}{8} +\frac{3S_{11}}{8} ,\\
P_y\left(-\frac{1}{2}\right) &=\frac{1}{4}-\frac{1}{10} S_T^y+\frac{1}{8} S_{L L}+\frac{1}{8} S_{T T}^{x x}-\frac{3}{8} S_{L L T}^y-\frac{1}{8} S_{T T T}^{y x x}=\frac{1}{4}-\frac{S_3}{10} +\frac{S_4}{8} +\frac{S_7}{8} -\frac{3S_{11}}{8} ,\\
P_y\left(-\frac{3}{2}\right) &=\frac{1}{4}-\frac{3}{10} S_T^y-\frac{1}{8} S_{L L}-\frac{1}{8} S_{T T}^{x x}+\frac{1}{8} S_{L L T}^y+\frac{1}{24} S_{T T T}^{y x x}=\frac{1}{4}-\frac{3S_3}{10} -\frac{S_4}{8} -\frac{S_7}{8} +\frac{S_{11}}{8} .
\end{aligned}
\end{equation}

In the intermediate expressions of Eq.~\eqref{xyz_probability}, we have included the complete results, which are valid for calculations in any process. However, in the  $e^{+}e^{-}\to \psi(3686)\to \Omega^{-}\bar{\Omega}^{+}$ process, certain polarization components, such as $S_L$ and $S_{LLL}$, are found to be zero. Therefore, in the final expression on the right-hand side, we have omitted these zero polarization components. Then, the values of these probabilities are determined exclusively by the non-zero parameters in Eq.~\eqref{coefficient_S}. The ranges of the aforementioned spin components,  as dictated by their corresponding physical interpretations in Appendix~\ref{A:Probabilistic}, are presented in Eq.~\eqref{spin_components_value}. 

According to Eq.~\eqref{xyz_probability}, it becomes evident that the influence of $S_{LL}$ extends beyond the $z$-axis and encompasses the polarization along the $x$- and $y$-axes as well. This peculiar polarization dependence observed in spin-3/2 particles contradicts the straightforward behavior exhibited by spin-1/2 particles. In essence, when the $\Omega^-$ particle possesses a non-zero $S_{LL}$ component, it not only induces disparities in polarization probabilities along the $z$-axis ($P_z(\pm\frac{3}{2})\neq P_z(\pm\frac{1}{2})$) but also gives rise to deviations in polarization probabilities along the $x$- and $y$-axes ($P_{x/y}(\pm\frac{3}{2})\neq P_{x/y}(\pm\frac{1}{2})$). The complexity arises from the intrinsic nature of spin-3/2, where the observation of spin in any direction can yield four distinct states. Investigating alternative forms of spin representation may prove worthwhile in mitigating this complexity. Nevertheless, the physical implications conveyed by Eq.~\eqref{xyz_probability} underscore the substantial advantages of projecting the spin of $\Omega^-$ particles along the coordinate axes. This approach simplifies the intricate polarization dependence of the $\Omega^-$ particle, facilitating a more intuitive understanding of its spin structure.

\begin{figure}
\centering
    \includegraphics[width=0.28\textwidth]{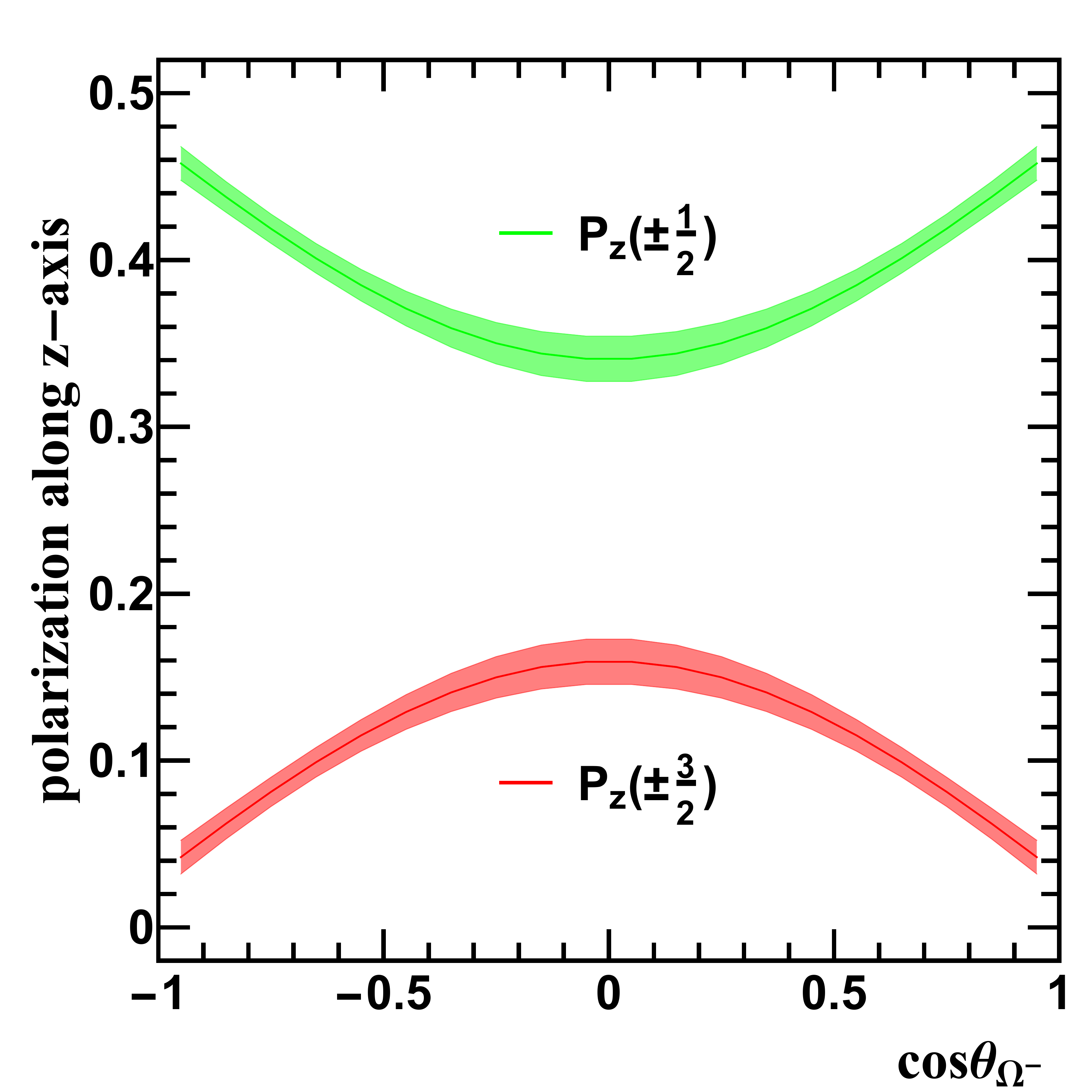}   \hspace{0.2cm}
   \includegraphics[width=0.28\textwidth]{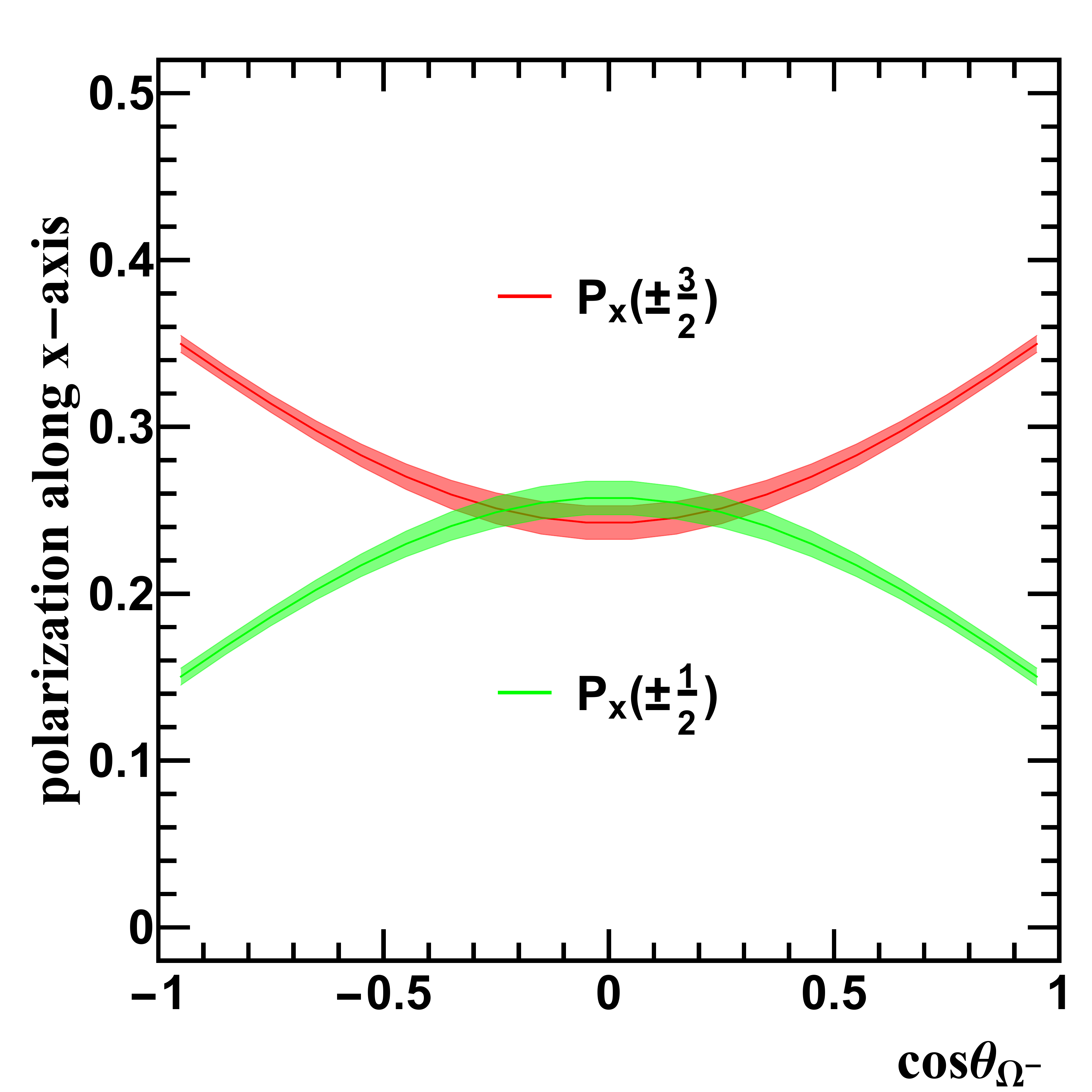}   \hspace{0.2cm}
   \includegraphics[width=0.28\textwidth]{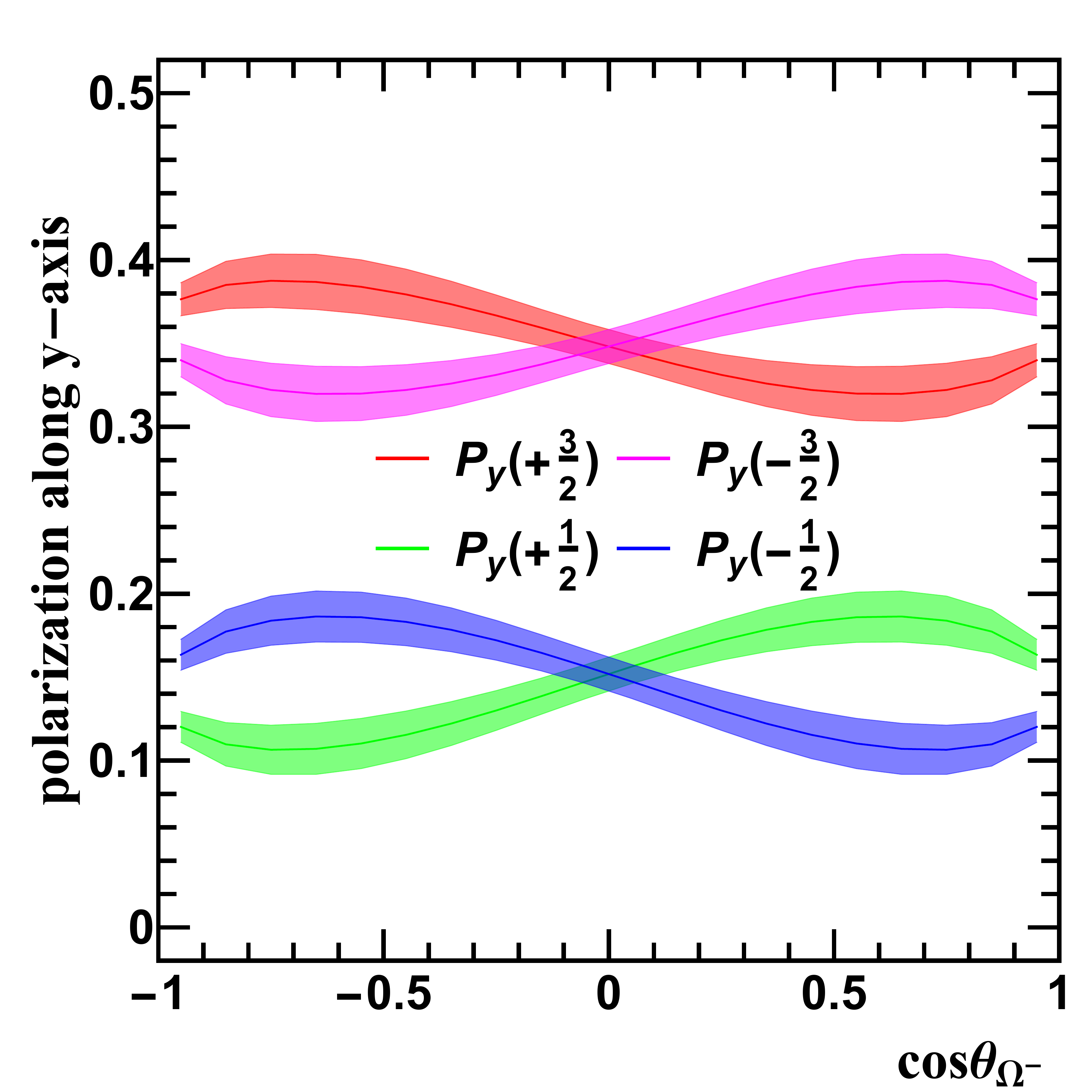}
\caption{The cross-section displays polarization dependence along the $z$-axis (left), $x$-axis (middle), and $y$-axis (right) as a function of the azimuthal angle $\cos\theta_{\Omega^{-}}$. The solid lines represent the central values, and the shaded areas
represent $\pm$ one standard deviation.}
\label{probability_Omega}
\end{figure}

By employing the computed multipolar polarization operators as described in Eq.~\eqref{coefficient_S}, we present a graph (Fig.~\ref{probability_Omega}) that illustrates the polarization dependence of the cross-section along the $x$, $y$, and $z$-axes with respect to $\cos\theta_{\Omega^{-}}$. All calculations were performed utilizing the helicity amplitudes presented in Table~\ref{simultaneous_result}. From Fig.~\ref{probability_Omega}, it is evident that the $\Omega^{-}$ particle exhibits significant tensor polarization along the $z$-axis in the $\psi(3686)\rightarrow\Omega^{-}\bar{\Omega}^{+}$ process. This is due to a substantial difference in the probability of occupying eigenstates with eigenvalues of $\pm\frac{3}{2}$ and $\pm\frac{1}{2}$ along the $z$-axis. Additionally, it demonstrates tensor polarization along the $x$-axis, which gradually increases in the direction of the electron beam and reaches its maximum at $\theta_{\Omega^{-}}=0$ or $\pi$. Furthermore, notable tensor polarization is observed along the $y$-axis, accompanied by weak transverse vector polarization and rank-3 tensor polarization. This is attributed to a significant difference in the probability of occupying eigenstates with eigenvalues of $\pm\frac{3}{2}$ and $\pm\frac{1}{2}$, while only a minor difference is observed in the probability of occupying eigenstates with eigenvalues of $\frac{3}{2}$ and $-\frac{3}{2}$ or $\frac{1}{2}$ and $-\frac{1}{2}$.

\subsection{cross-section for $e^{+}e^{-}\rightarrow\Omega^{-}\bar{\Omega}^{+}$} \label{s.cross}

\begin{figure}[ht]
    \centering %
\includegraphics[width=0.50\textwidth]{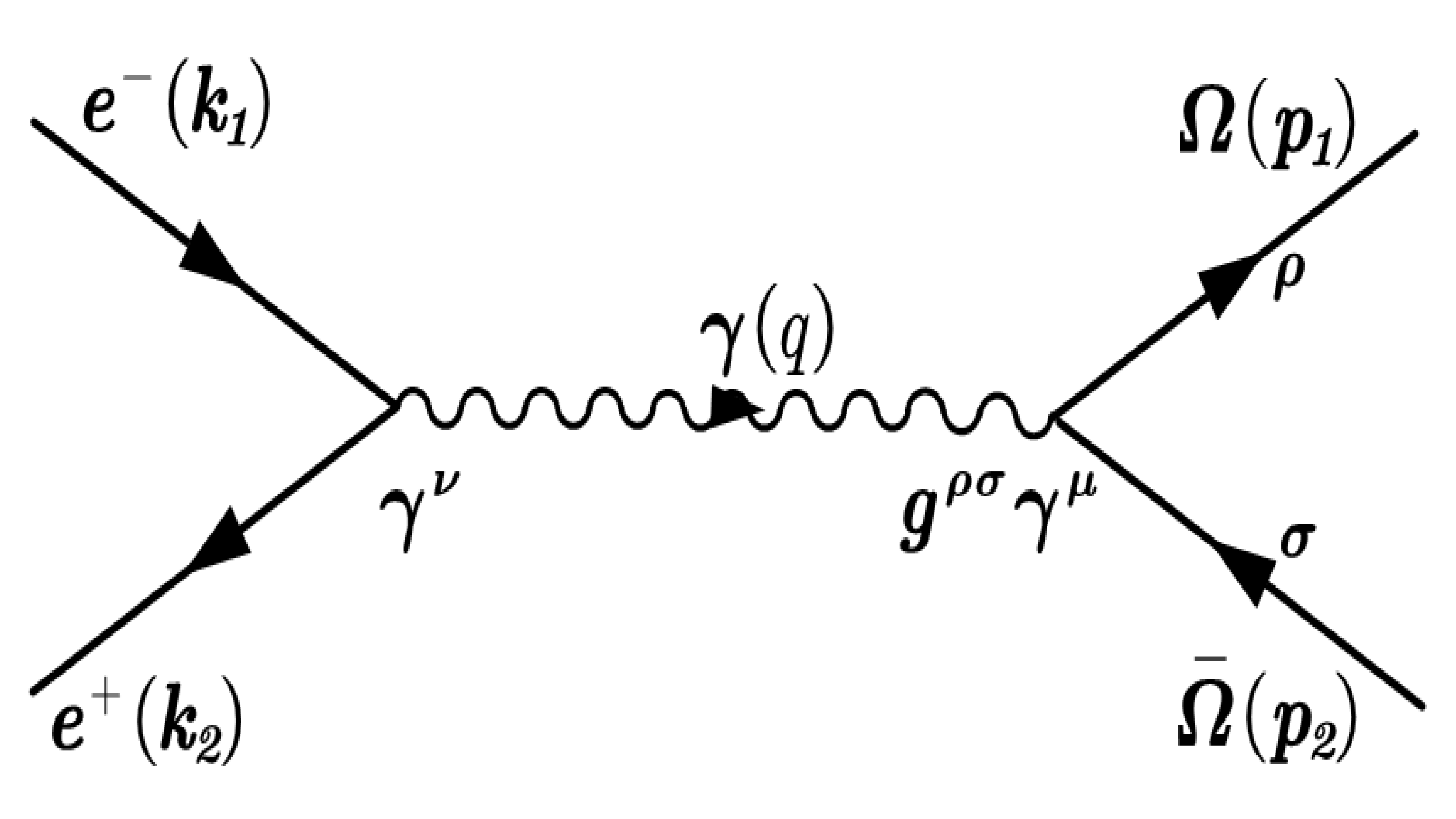}

\caption{Graph describing the reaction $e^{+}e^{-}\rightarrow\Omega^{-}\bar{\Omega}^{+}$
in one-photon approximation.}
\label{f.Graph}
\end{figure}

In order to provide a theoretical interpretation for the experimental measurements, we calculate the polarization dependence of the cross-section in the process of $e^+e^-$ annihilation into $\Omega^{-}\bar{\Omega}^{+}$ using a point-like vertex, as depicted in Fig.~\ref{f.Graph}. Since both the photon and $\psi(3686)$ are vector particles, we neglect the influence of the vector boson $\psi(3686)$ on the cross-section. The calculations are performed in the center-of-mass frame of the leptons, as shown in Fig.~\ref{helicity_angle}. Neglecting the electronic mass, the momentum components of $e^+e^-$ and $\Omega^{-}\bar{\Omega}^{+}$ can be expressed as follows:
\begin{align}
k_{1} & =  \left\{ E,0,0,-E\right\} ,\\
k_{2} & =  \left\{ E,0,0,E\right\} ,\\
p_{1} & =  \left\{ E,\left|\vec{p}\right|\sin\theta_{\Omega},0,\left|\vec{p}\right|\cos\theta_{\Omega}\right\} ,\\
p_{2} & =  \left\{ E,-\left|\vec{p}\right|\sin\theta_{\Omega},0,-\left|\vec{p}\right|\cos\theta_{\Omega}\right\} .
\end{align}

In the $e^+e^- \rightarrow \Omega^- \bar{\Omega}^+$ process, the cross-section can be expressed in terms of the scattering amplitude. When considering the polarization dependence of the cross-section along the helicity direction, we project the polarizations of the initial and final state particles onto their respective helicity directions. The expression for the cross-section is given by:
\begin{align}
\frac{d\sigma^{\lambda_{1}}}{d\Omega}  =  \frac{1}{64\pi^{2}s}\frac{\left|\vec{p}\right|}{E}\frac{1}{4}\sum_{\lambda=\pm}\sum_{\lambda_{2}=\pm\frac{1}{2},\pm\frac{3}{2}}\left|\mathcal{M}_{-\lambda,\lambda,\lambda_{1},\lambda_{2}}\right|^{2},
\end{align}
where $\lambda$, $\lambda_{1}$, and $\lambda_{2}$ represent the helicities of the positron, $\Omega^{-}$, and $\bar{\Omega}^{+}$, respectively. The scattering amplitude $\mathcal{M}$ can be expressed as follows:
\begin{align}
\mathcal{M}_{-\lambda,\lambda,\lambda_{1},\lambda_{2}}  =  \frac{g^{\mu\nu}g^{\rho\sigma}}{q^{2}}\left[\bar{u}^{\rho}\left(p_{1},\lambda_{1}\right)\left(-ie\gamma^{\mu}\right)v^{\sigma}\left(p_{2},\lambda_{2}\right)\right]\left[\bar{v}\left(k_{2},\lambda\right)\left(-ie\gamma^{\nu}\right)u\left(k_{1},-\lambda\right)\right],
\end{align}
where $\bar{u}^{\rho}(v^{\sigma})$ and $u(\bar{v})$ represent the polarization vectors for $\Omega^{-}$ ($\bar{\Omega}^{+}$) and $e^{-}$ ($e^{+}$) respectively, following the conventions described in Appendix~\ref{C:Polarization vectors}. When investigating the dependence of the cross-section on the transverse polarization of $\Omega^-$, it is essential to decompose its spin within its own helicity frame, specifically along the directions $\hat{x}_1$ or $\hat{y}_1$, as illustrated in Fig.~\ref{helicity_angle}. However, for the other particles, their polarization states are not of interest, and thus, we sum over their spins. It is important to note that any spin decomposition for these particles along different directions does not affect the final result. Therefore, we continue to employ their spin decomposition along their own helicity direction. We can now express the cross-section's dependence on the transverse polarization of $\Omega^-$ as follows:
\begin{align}
\frac{d\sigma^{s_{1}\uparrow_{x/y}}}{d\Omega} = \frac{1}{64\pi^{2}s}\frac{\left|\vec{p}\right|}{E}\frac{1}{4}\sum_{\lambda=\pm}\sum_{\lambda_{2}=\pm\frac{1}{2},\pm\frac{3}{2}}\left|\mathcal{M}_{-\lambda,\lambda,s_{1}\uparrow_{x/y},\lambda_{2}}\right|^{2},
\end{align}
where we introduce the notation $s_{1}\uparrow_{x/y}$ to represent the transverse polarization state of the $\Omega^{-}$ particle. The expressions for all the scattering amplitudes are provided in Appendix~\ref{D:scattering amplitudes}. Based on the aforementioned considerations, we derive the polarization dependence of the cross-section as follows:
\begin{align}
\frac{d\sigma^{\pm\frac{3}{2}}}{d\Omega}  = & \frac{\alpha^{2}}{32E^{2}}\sqrt{1-\frac{m^{2}}{E^{2}}}\left\{ \left(a^{2}+\frac{m^{2}}{E^{2}}\right)+\left(a^{2}-\frac{m^{2}}{E^{2}}\right)\cos^{2}\theta_{\Omega}\right\} ,\label{e.z3}\\
\frac{d\sigma^{\pm\frac{1}{2}}}{d\Omega}  = & \frac{\alpha^{2}}{32E^{2}}\sqrt{1-\frac{m^{2}}{E^{2}}}\left\{ \left[a^{2}+b^{4}\left(\frac{2E^{2}}{m^{2}}-1\right)^{2}+\left(\frac{2E}{m}b^{2}-\frac{m}{E}\left(a^{2}+b^{2}\right)\right)^{2}\right]\right.\nonumber\\
 & \qquad\left.+\left[a^{2}+b^{4}\left(\frac{2E^{2}}{m^{2}}-1\right)^{2}-\left(\frac{2E}{m}b^{2}-\frac{m}{E}\left(a^{2}+b^{2}\right)\right)^{2}\right]\cos^{2}\theta_{\Omega}\right\} \label{e.z1},\\
\frac{d\sigma^{\pm\frac{3}{2}\uparrow_{x}}}{d\Omega}  = &\frac{\alpha^{2}}{32E^{2}}\sqrt{1-\frac{m^{2}}{E^{2}}}\left\{ \left[a^{2}\left(1-3b^{2}\right)+\frac{\sqrt{3}}{2}ab^{2}-\frac{9}{4}b^{4}\right.\right.\nonumber\\
 & \qquad\left.+\frac{3E^{4}}{m^{4}}b^{4}-\frac{\sqrt{3}E^{2}}{m^{2}}ab^{2}+\frac{m^{2}}{4E^{2}}\left(1+3\left(a^{2}+b^{2}\right)^{2}\right)\right]\nonumber\\
 & \quad+\left[a^{2}\left(1+3b^{2}\right)-\frac{\sqrt{3}}{2}ab^{2}+\frac{15}{4}b^{4}+\frac{3E^{4}}{m^{4}}b^{4}\right.\nonumber\\
 & \qquad\left.\left.+\frac{E^{2}}{m^{2}}b^{2}\left(\sqrt{3}a-6b^{2}\right)-\frac{m^{2}}{4E^{2}}\left(1+3\left(a^{2}+b^{2}\right)^{2}\right)\right]\cos^{2}\theta_{\Omega}\right\} \label{e.x3},
\end{align}
\begin{align}
\frac{d\sigma^{\pm\frac{1}{2}\uparrow_{x}}}{d\Omega}  = &\frac{\alpha^{2}}{32E^{2}}\sqrt{1-\frac{m^{2}}{E^{2}}}\left\{ \left[a^{2}\left(1-b^{2}\right)-\frac{\sqrt{3}}{2}ab^{2}-\frac{3}{4}b^{4}\right.\right.\nonumber\\
 & \qquad\left.+\frac{E^{4}}{m^{4}}b^{4}+\frac{\sqrt{3}E^{2}}{m^{2}}ab^{2}+\frac{m^{2}}{4E^{2}}\left(3+\left(a^{2}+b^{2}\right)^{2}\right)\right]\nonumber\\
 & \quad+\left[a^{2}\left(1+b^{2}\right)+\frac{\sqrt{3}}{2}ab^{2}+\frac{5}{4}b^{4}+\frac{E^{4}}{m^{4}}b^{4}\right.\nonumber\\
 & \qquad\left.\left.-\frac{E^{2}}{m^{2}}b^{2}\left(\sqrt{3}a+2b^{2}\right)-\frac{m^{2}}{4E^{2}}\left(3+\left(a^{2}+b^{2}\right)^{2}\right)\right]\cos^{2}\theta_{\Omega}\right\} \label{e.x1},\\
 \frac{d\sigma^{\pm\frac{3}{2}\uparrow_{y}}}{d\Omega} & =\frac{\alpha^{2}}{32E^{2}}\sqrt{1-\frac{m^{2}}{E^{2}}}\left\{ \left[a^{2}\left(1-3b^{2}\right)-\frac{\sqrt{3}}{2}ab^{2}-\frac{9}{4}b^{4}\right.\right.\nonumber\\
 & \qquad\left.+\frac{3E^{4}}{m^{4}}b^{4}+\frac{\sqrt{3}E^{2}}{m^{2}}ab^{2}+\frac{m^{2}}{4E^{2}}\left(1+3\left(a^{2}+b^{2}\right)^{2}\right)\right]\nonumber\\
 & \quad+\left[a^{2}\left(1+3b^{2}\right)+\frac{\sqrt{3}}{2}ab^{2}+\frac{15}{4}b^{4}+\frac{3E^{4}}{m^{4}}b^{4}\right.\nonumber\\
 & \qquad\left.\left.-\frac{E^{2}}{m^{2}}b^{2}\left(\sqrt{3}a+6b^{2}\right)-\frac{m^{2}}{4E^{2}}\left(1+3\left(a^{2}+b^{2}\right)^{2}\right)\right]\cos^{2}\theta_{\Omega}\right\} \label{e.y3},\\
  \frac{d\sigma^{\pm\frac{1}{2}\uparrow_{y}}}{d\Omega}  = & \frac{\alpha^{2}}{32E^{2}}\sqrt{1-\frac{m^{2}}{E^{2}}}\left\{ \left[a^{2}\left(1-b^{2}\right)+\frac{\sqrt{3}}{2}ab^{2}-\frac{3}{4}b^{4}\right.\right.\nonumber\\
 & \qquad\left.+\frac{E^{4}}{m^{4}}b^{4}-\frac{\sqrt{3}E^{2}}{m^{2}}ab^{2}+\frac{m^{2}}{4E^{2}}\left(3+\left(a^{2}+b^{2}\right)^{2}\right)\right]\nonumber\\
 & \quad+\left[a^{2}\left(1+b^{2}\right)-\frac{\sqrt{3}}{2}ab^{2}+\frac{5}{4}b^{4}+\frac{E^{4}}{m^{4}}b^{4}\right.\nonumber\\
 & \qquad\left.\left.+\frac{E^{2}}{m^{2}}b^{2}\left(\sqrt{3}a-2b^{2}\right)-\frac{m^{2}}{4E^{2}}\left(3+\left(a^{2}+b^{2}\right)^{2}\right)\right]\cos^{2}\theta_{\Omega}\right\} \label{e.y1},
 \end{align}
where $m$ denotes the mass of the $\Omega^-$ particle, $a$ and $b$ are normalization coefficients satisfying $a^2 + b^2 = 1$. In the SU(6) quark model, one can determine $a = \sqrt{1/3}$ and $b = \sqrt{2/3}$, as presented in Appendix~\ref{C:Polarization vectors}. We retain $a$ and $b$ as coefficients and demonstrate the dependence of the cross-section on $a$ and $b$. It is important to note that altering the values of $a$ and $b$ introduces certain challenges, which will be elaborated upon in the subsequent paragraph.

\begin{figure}[ht]
    \centering %
    \includegraphics[width=0.28\textwidth]{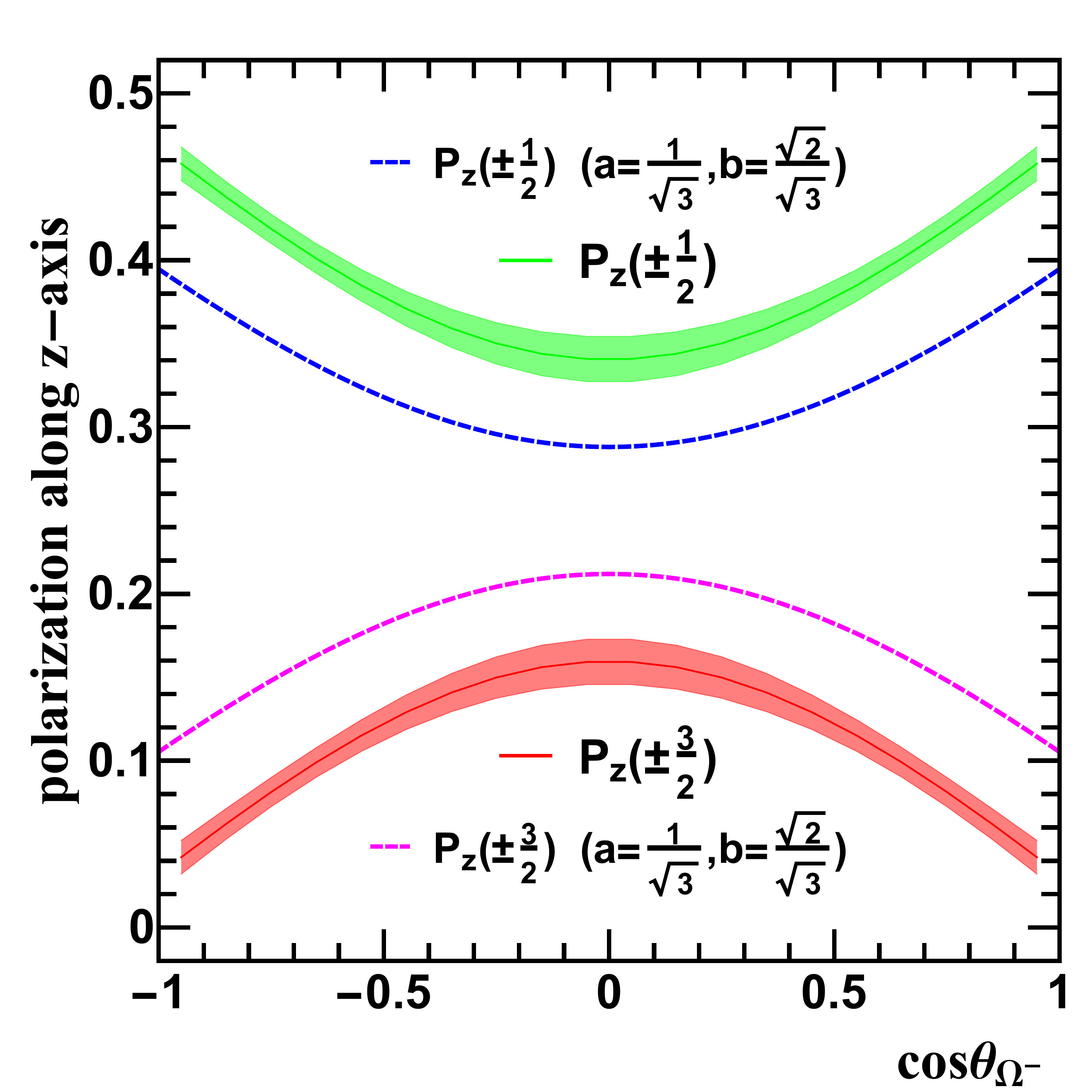}
    \hspace{0.2cm}
    \includegraphics[width=0.28\textwidth]{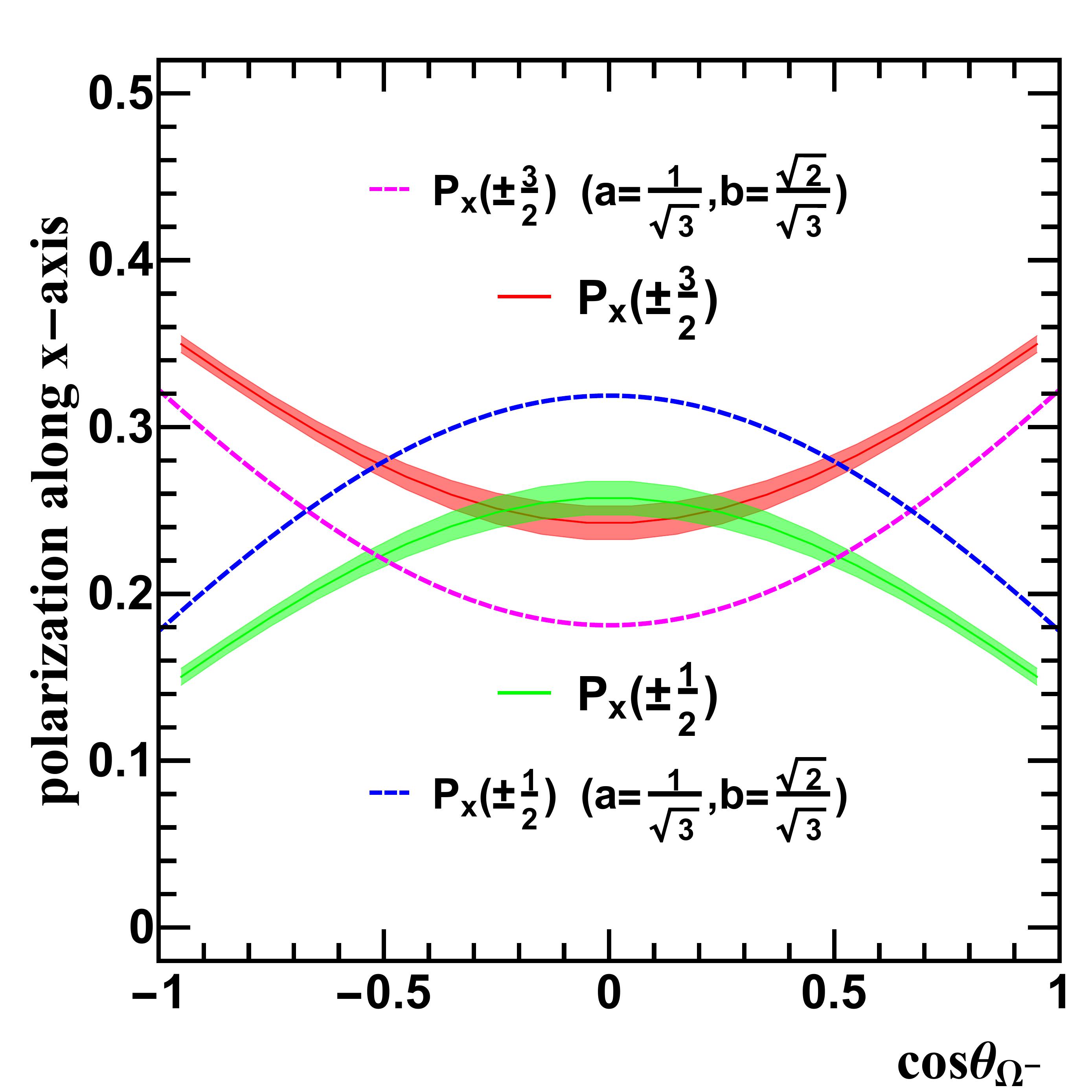}
        \hspace{0.2cm}
    \includegraphics[width=0.28\textwidth]{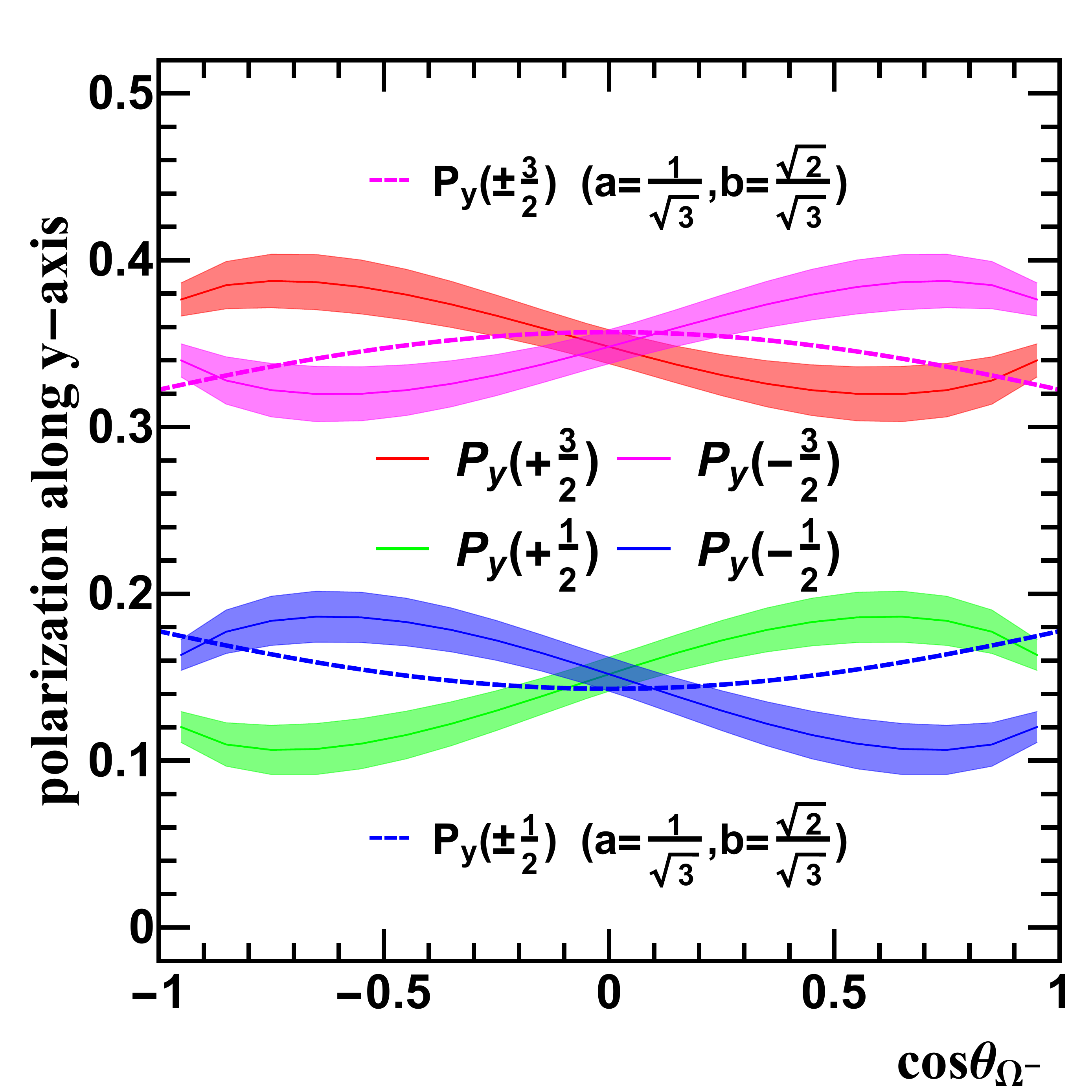}\\
    \centering %
    \includegraphics[width=0.28\textwidth]{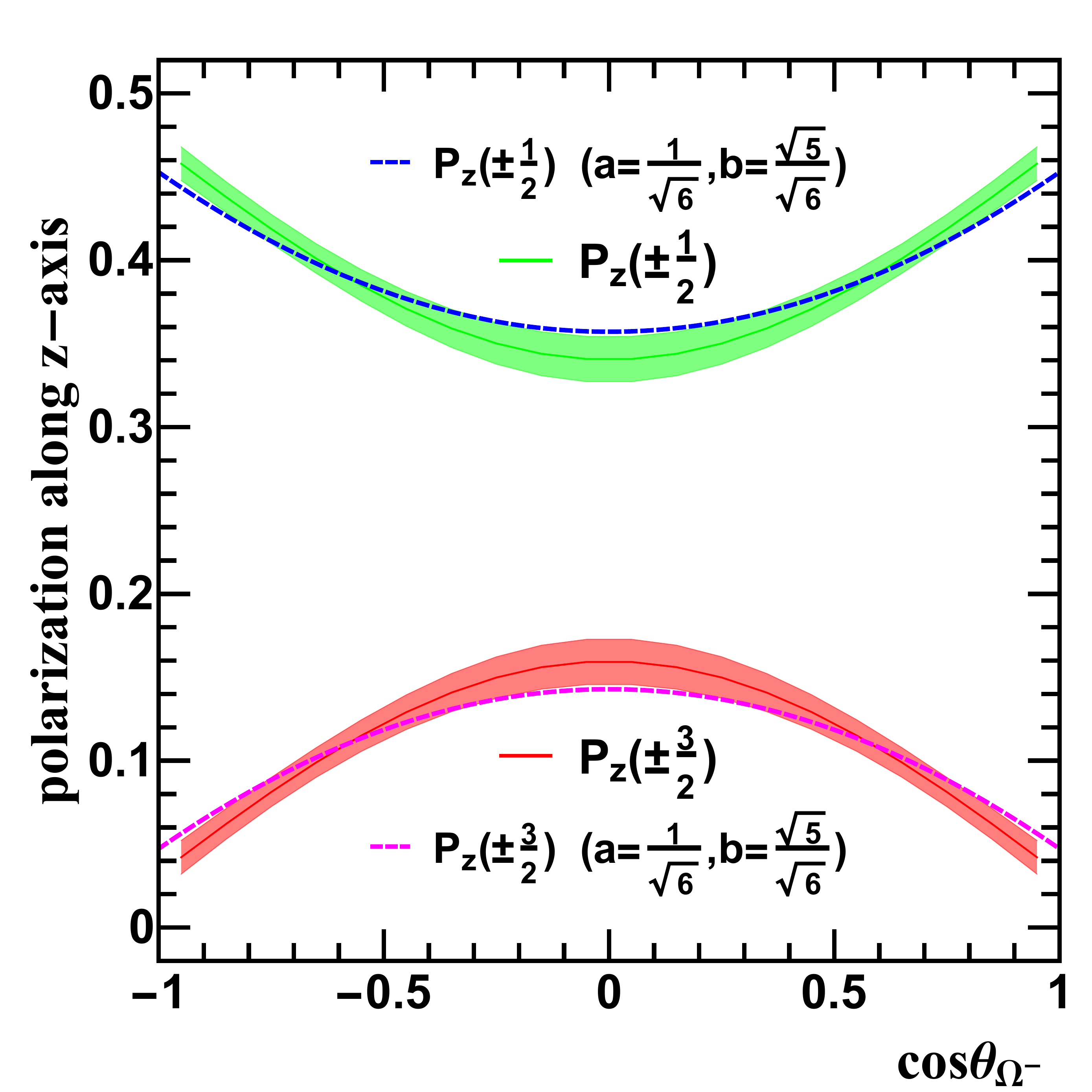}
    \hspace{0.2cm}
    \includegraphics[width=0.28\textwidth]{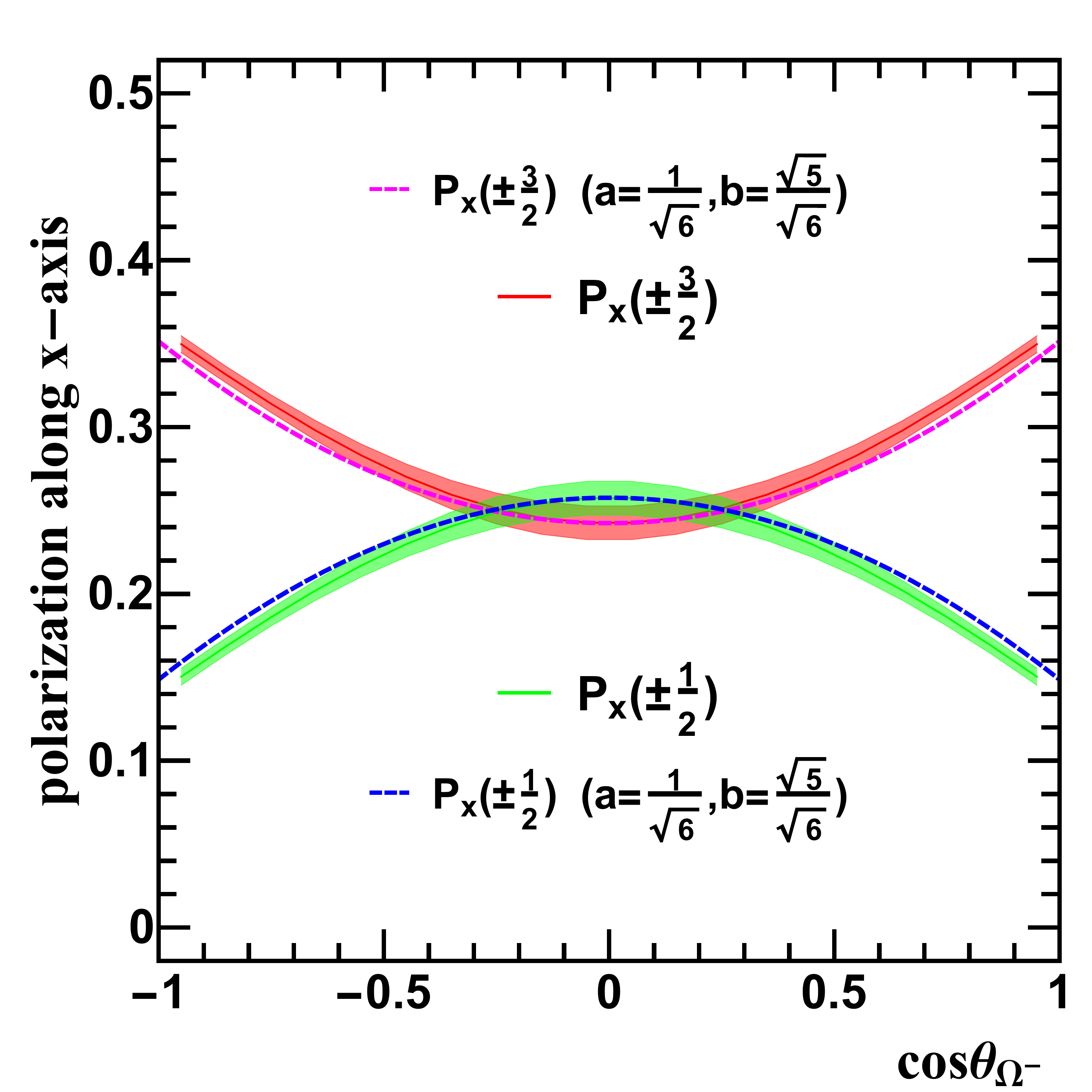}
        \hspace{0.2cm}
    \includegraphics[width=0.28\textwidth]{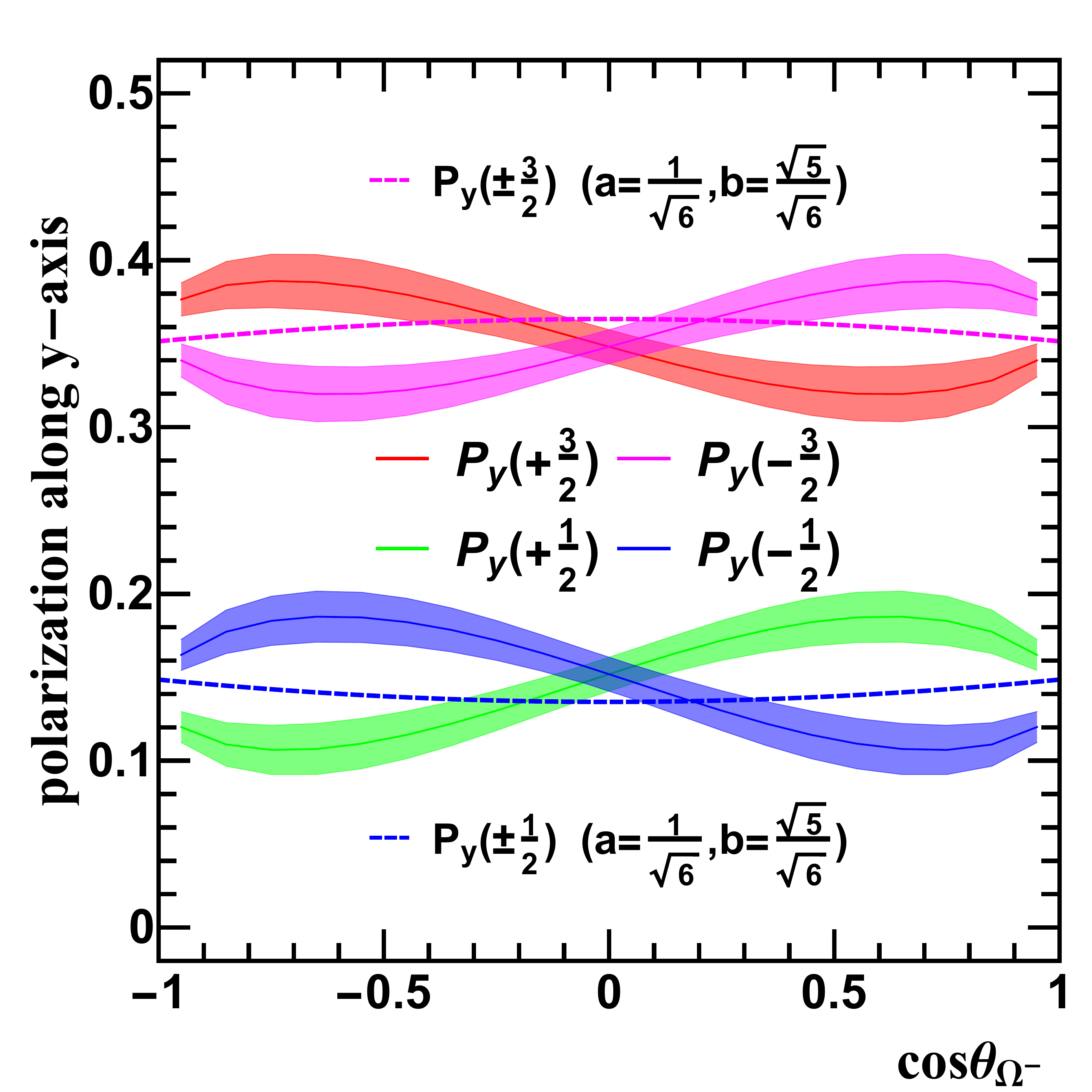}
   \caption{The dashed lines represent the polarization dependence for $e^{+}e^{-}\rightarrow\Omega^{-}\bar{\Omega}^{+}$ at $\sqrt{s}=3.686$ GeV, calculated theoretically using Eqs.~\eqref{e.z3}--\eqref{e.y1}. The solid lines correspond to the results obtained from the BESIII experiment\cite{BESIII:2020lkm}.}
\label{f.theoty}
\end{figure}

In Fig.~\ref{f.theoty}, we present the theoretical predictions for the polarization dependence of the cross-section in the $e^{+}e^{-}\rightarrow\Omega^{-}\bar{\Omega}^{+}$ process. The corresponding experimental measurements are also included for comparison. Based on the graphical representation, it is apparent that the azimuth-angle dependence of the polarization cross-section agrees with the experimental measurements when $a=\sqrt{1/3}$ and $b=\sqrt{2/3}$. Nevertheless, when we allow the coefficients $a$ and $b$ to vary, it leads to a corresponding variation in the polarization dependence of the cross-section. Remarkably, when $a=\sqrt{1/6}$ and $b=\sqrt{5/6}$ are chosen, our calculations exhibit a notably improved agreement with the experimental measurements reported in Ref.~\cite{BESIII:2020lkm}. This observation suggests that the field equation of $\Omega^{-}$ may deviate from the predictions of the SU(6) quark model. Additionally, the coefficients $a$ and $b$ play a significant role in eliminating the spin-1/2 component from the field equation, indicating that $\Omega^{-}$ may not exclusively exist in a pure spin-3/2 state. This intriguing finding offers novel insights for future investigations and extensive discussions regarding the nature and potential dynamics of $\Omega^{-}$. It is important to acknowledge that our calculations did not consider the influence of intermediate resonance states ($\psi(3686)$) and form factors of $\Omega^{-}$ on the cross-section. The incorporation of these factors presents intriguing prospects for future research and additional exploration.

\section{Summary} \label{S:summary}
In this paper, we investigated the polarization dependence of the cross-section for the production of $\Omega^{-}$ particles in the $e^{+}e^{-}\rightarrow\psi(3686)\rightarrow\Omega^{-}\bar{\Omega}^{+}$ process.

We conducted a review of the application of the helicity formalism analysis method to the $e^{+}e^{-}\rightarrow\psi(3686)\rightarrow\Omega^{-}\bar{\Omega}^{+}$ process. In this analysis, the cross-section can be described by four independent helicity transition amplitudes: $H_1$, $H_2$, $H_3$, and $H_4$, which are associated with the spin density matrix of  $\Omega^{-}$ and $\bar{\Omega}^{+}$. The polarization state of spin-3/2 particles can be characterized by 15 independent polarization components. However, it is important to note that the decomposition of these polarization components is not consistent among different articles. To address this issue, we propose a redefinition of the spin matrix basis in the helicity formalism. This redefinition establishes a direct correspondence between the spin components used in the helicity formalism and those defined by the spin density matrix. In the $e^{+}e^{-}\rightarrow\psi(3686)\rightarrow\Omega^{-}\bar{\Omega}^{+}$ process, we observe that the $\Omega^{-}$ particle exhibits six non-zero polarization states, namely $S_3$ ($S_T^{y}$), $S_4$ ($S_{LL}$), $S_5$ ($S_{LT}^{x}$), $S_7$ ($S_{TT}^{xx}$), $S_{11}$ ($S_{LLT}^{y}$), and $S_{13}$ ($S_{LTT}^{xy}$). By utilizing experimental measurements obtained from BESIII, we present plots that illustrate the variation of the polarization-dependent reaction cross-section as a function of the azimuthal angle.

To enhance our understanding of the polarization states of the $\Omega^-$ particle, we conduct an analysis of the polarization dependence of the reaction cross-section by projecting it onto the Cartesian coordinate axes. This analysis provides valuable insights into the contributions of different polarization components along the $x$-, $y$-, and $z$-axes, elucidating their individual impacts on the overall cross-section. Our results demonstrate significant tensor polarization of the $\Omega^{-}$ particle along the $z$-axis and $y$-axis. Moreover, there is a notable tensor polarization along the $x$-axis, particularly when $\theta_{\Omega}$ takes values of zero or $\pi$. However, we observe only a weak transverse vector polarization along the $y$-axis.

To account for the polarization dependence of the reaction cross-section, we conduct calculations for the cross-section distribution in the $e^{+}e^{-}\rightarrow\Omega^{-}\bar{\Omega}^{+}$ process. Our calculations do not incorporate the form factors of the $\Omega^{-}$ particles or the influence of intermediate resonance states($\psi(3686)$). However, despite these simplifications, our calculations exhibit a similar trend to the experimental measurements. Moreover, we observe that the properties of the $\Omega^{-}$ particle deviate to some extent from the predictions of the SU(6) quark model, suggesting that the field associated with the $\Omega^{-}$ particle impacts the underlying coefficients.

\section{Acknowledgements} \label{A:ackn}

 The authors would like to thank Prof. Zuo-tang Liang, and Prof. Tianbo  Liu for useful discussions and and suggestions. This work was supported by the National Natural Science Foundation of China (Grant No. 12247121).

\newpage

\appendix

\section{Probabilistic interpretation of the spin component}\label{A:Probabilistic}

To present the explicit expression of the probabilistic interpretations of spin polarization using Eq.~\eqref{f:sigmaij} and Eq.~\eqref{f:sigmaijk}, we begin by decomposing the spin operators for spin-3/2 particles into the following forms:
\begin{equation}\label{simplification}
\begin{aligned}
&\Sigma^{xz}=\frac{1}{4}(\Sigma^{x}+\Sigma^{z})^{2}-\frac{1}{4}(\Sigma^{x}-\Sigma^{z})^{2},\\
&\Sigma^{yz}=\frac{1}{4}(\Sigma^{y}+\Sigma^{z})^{2}-\frac{1}{4}(\Sigma^{y}-\Sigma^{z})^{2},\\
&\Sigma^{zz}=(\Sigma^{z})^2- \frac{5}{4},\\
&\Sigma^{xy}=\frac{1}{4}(\Sigma^{x}+\Sigma^{y})^{2}-\frac{1}{4}(\Sigma^{x}-\Sigma^{y})^{2},\\
&\Sigma^{xx}-\Sigma^{yy}=(\Sigma^{x})^{2}-(\Sigma^{y})^{2},\\
&\Sigma^{xzz}=\frac{1}{3}[\frac{1}{2}(\Sigma^{x}+\Sigma^{z})^{3}+\frac{1}{2}(\Sigma^{x}-\Sigma^{z})^{3}-(\Sigma^{x})^{3}]-\frac{41}{60}\Sigma^{x},\\
&\Sigma^{yzz}=\frac{1}{3}[\frac{1}{2}(\Sigma^{y}+\Sigma^{z})^{3}+\frac{1}{2}(\Sigma^{y}-\Sigma^{z})^{3}-(\Sigma^{y})^{3}]-\frac{41}{60}\Sigma^{y},\\
&\Sigma^{zzz}=(\Sigma^{z})^3- \frac{41}{20}\Sigma^{z},\\
&\Sigma^{xyz}=\frac{1}{6}[(\Sigma^{x}+\Sigma^{y}+\Sigma^{z})^{3}-(\Sigma^{x}+\Sigma^{y})^{3}-(\Sigma^{x}+\Sigma^{z})^{3}-(\Sigma^{y}+\Sigma^{z})^{3}+(\Sigma^{x})^{3}+(\Sigma^{y})^{3}+(\Sigma^{z})^{3}],\\
&\Sigma^{xxz}-\Sigma^{yyz}=\frac{1}{6}[(\Sigma^{x}+\Sigma^{z})^{3}-(\Sigma^{x}-\Sigma^{z})^{3}-(\Sigma^{y}+\Sigma^{z})^{3}+(\Sigma^{y}-\Sigma^{z})^{3}],\\
&\Sigma^{xxx}-3\Sigma^{xyy}=2(\Sigma^{x})^{3}-\frac{1}{2}(\Sigma^{x}+\Sigma^{y})^{3}-\frac{1}{2}(\Sigma^{x}-\Sigma^{y})^{3},\\
&3\Sigma^{yxx}-\Sigma^{yyy}=-2(\Sigma^{y})^{3}+\frac{1}{2}(\Sigma^{y}+\Sigma^{x})^{3}+\frac{1}{2}(\Sigma^{y}-\Sigma^{x})^{3}.\\
\end{aligned}
\end{equation}
 Within this framework, we define the eigenstates $|m_{(\theta,\phi)}\rangle$ associated with the spin projection operator along the direction $(\theta,\phi)$. Here, $\phi$ represents the azimuthal angle, and $\theta$ represents the polar angle. These eigenstates have corresponding eigenvalues denoted as $m$. In addition, for the sake of clarity and convenience, we introduce the following notation:
 \begin{equation}
 \begin{aligned}
|m\rangle_{x+y}&=\left|m_{(\frac{\pi}{2},\frac{\pi}{4})}\right\rangle, \quad|m\rangle_{x+z}=\left|m_{\left(\frac{\pi}{4}, 0\right)}\right\rangle, \quad|m\rangle_{y+z}=\left|m_{\left(\frac{\pi}{4}, \frac{\pi}{2}\right)}\right\rangle,\\
|m\rangle_{x-y}&=\left|m_{(\frac{\pi}{2},-\frac{\pi}{4})}\right\rangle, \quad|m\rangle_{x-z}=\left|m_{\left(-\frac{\pi}{4}, 0\right)}\right\rangle, \quad|m\rangle_{y-z}=\left|m_{\left(-\frac{\pi}{4}, \frac{\pi}{2}\right)}\right\rangle,\\
|m\rangle_{x+y+z}&=\left|m_{(\theta_{xyz},\frac{\pi}{4})}\right\rangle,\\
\end{aligned}
\end{equation}
where $\theta_{xyz}$=arctan($\sqrt{2}$).

The probability of the system being in a particular eigenstate can be determined by evaluating the spin density matrix using the formula given by Eq.~\eqref{d_probability}. Subsequently, all the spin components can be expressed in terms of the probabilities $P(m_{(\theta,\phi)})$s. The three spin components of the spin vector $S^{i}$ are given by~\cite{Zhao:2022lbw}:
\begin{equation}\label{coefficient_1}
\begin{aligned}
 S_{L}&=\frac{3}{2}[P_{z}(\frac{3}{2})-P_{z}(-\frac{3}{2})]+\frac{1}{2}[P_{z}(\frac{1}{2})-P_{z}(-\frac{1}{2})],
\end{aligned}
\end{equation}
\begin{equation}\label{coefficient_2}
\begin{aligned}
  S_{T}^{x}&=
 \frac{3}{2}[P_{x}(\frac{3}{2})-P_{x}(-\frac{3}{2})]+
 \frac{1}{2}[P_{x}(\frac{1}{2})-P_{x}(-\frac{3}{2})],
\end{aligned}
\end{equation}
\begin{equation}\label{coefficient_3}
\begin{aligned}
 S_{T}^{y}&=
 \frac{3}{2}[P_{y}(\frac{3}{2})-P_{y}(-\frac{3}{2})]+
 \frac{1}{2}[P_{y}(\frac{1}{2})-P_{y}(-\frac{1}{2})].
\end{aligned}
\end{equation}
The five spin components of the rank-2 spin tensor $T^{ij}$ are given by~\cite{Zhao:2022lbw}:
\begin{equation}\label{coefficient_4}
\begin{aligned}
S_{LL} &= [P_{z}(\frac{3}{2})+P_{z}(-\frac{3}{2})]-[P_{z}(\frac{1}{2})+P_{z}(-\frac{1}{2})],
\end{aligned}
\end{equation}
\begin{equation}\label{coefficient_5}
\begin{aligned}
  S^{x}_{LT}&=
  2[P_{x+z}(\frac{3}{2})+P_{x+z}(-\frac{3}{2})]-
  2[P_{x-z}(\frac{3}{2})+P_{x-z}(-\frac{3}{2})],
\end{aligned}
\end{equation}
\begin{equation}\label{coefficient_6}
\begin{aligned}
 S^{y}_{LT} &=
2[P_{y+z}(\frac{3}{2})+P_{y+z}(-\frac{3}{2})]-
 2[P_{y-z}(\frac{3}{2})+P_{y-z}(-\frac{3}{2})],
\end{aligned}
\end{equation}
\begin{equation}\label{coefficient_7}
\begin{aligned}
S_{TT}^{xx}  &= 2[P_{x}(\frac{3}{2})+P_{x}(-\frac{3}{2})]-
2[P_{y}(\frac{3}{2})+P_{y}(-\frac{3}{2})],
\end{aligned}
\end{equation}
\begin{equation}\label{coefficient_8}
\begin{aligned}
S_{TT}^{xy}  &=
 2[P_{x+y}(\frac{3}{2})+P_{x+y}(-\frac{3}{2})]
 -2[P_{x-y}(\frac{3}{2})+P_{x-y}(-\frac{3}{2})].
\end{aligned}
\end{equation}

The seven spin components of the rank-3 spin tensor $R^{ijk}$ are given by~\cite{Zhao:2022lbw}:
\begin{equation}\label{coefficient_9}
\begin{aligned}
S_{LLL} &= \frac{3}{10}[P_{z}(\frac{3}{2})-P_{z}(-\frac{3}{2})]-\frac{9}{10}[P_{z}(\frac{1}{2})-P_{z}(-\frac{1}{2})],
\end{aligned}
\end{equation}
\begin{equation}\label{coefficient_10}
\begin{aligned}
S_{LLT}^{x} &=
\frac{\sqrt{2}}{24}\{27[P_{x+z}(\frac{3}{2})-P_{x+z}(-\frac{3}{2})]+
[P_{x+z}(\frac{1}{2})-P_{x+z}(-\frac{1}{2})]\}\\
 &+\frac{\sqrt{2}}{24}\{27[(P_{x-z}(\frac{3}{2})-P_{x-z}(-\frac{3}{2}))
 +[P_{x-z}(\frac{1}{2})-P_{x-z}(-\frac{1}{2})]\}\\
  &-\frac{1}{60}\{129[P_{x}(\frac{3}{2})-P_{x}(-\frac{3}{2})]
  +23[P_{x}(\frac{1}{2})-P_{x}(-\frac{1}{2})]\},
\end{aligned}
\end{equation}
\begin{equation}\label{coefficient_11}
\begin{aligned}
 S_{LLT}^{y} &=
 \frac{\sqrt{2}}{24}\{27[P_{y+z}(\frac{3}{2})-P_{y+z}(-\frac{3}{2})]+
[P_{y+z}(\frac{1}{2})-P_{y+z}(-\frac{1}{2})]\}\\
 &+\frac{\sqrt{2}}{24}\{27[(P_{y-z}(\frac{3}{2})-P_{y-z}(-\frac{3}{2}))
 +[P_{y-z}(\frac{1}{2})-P_{y-z}(-\frac{1}{2})]\}\\
  &-\frac{1}{60}\{129[P_{y}(\frac{3}{2})-P_{y}(-\frac{3}{2})]
  +23[P_{y}(\frac{1}{2})-P_{y}(-\frac{1}{2})]\},
\end{aligned}
\end{equation}
\begin{equation}\label{coefficient_12}
\begin{aligned}
 S_{LTT}^{xy} &=
 \frac{1}{12}\{27[P_{z}(\frac{3}{2})-P_{z}(-\frac{3}{2})]+[P_{z}(\frac{1}{2})-P_{z}(-\frac{1}{2})]\}\\
 &+\frac{1}{12}\{27[P_{x}(\frac{3}{2})-P_{x}(-\frac{3}{2})]
 +[P_{x}(\frac{1}{2})-P_{x}(-\frac{1}{2})]\}\\
 &+\frac{1}{12}\{27[P_{y}(\frac{3}{2})-P_{y}(-\frac{3}{2})]
 +[P_{y}(\frac{1}{2})-P_{y}(-\frac{1}{2})]\}\\
 &-\frac{\sqrt{2}}{6}\{27[P_{x+z}(\frac{3}{2})-P_{x+z}(-\frac{3}{2})]
 +[P_{x+z}(\frac{1}{2})-P_{x+z}(-\frac{1}{2})]\}\\
  &-\frac{\sqrt{2}}{6}\{27[P_{y+z}(\frac{3}{2})-P_{y+z}(-\frac{3}{2})]
  +[P_{y+z}(\frac{1}{2})-P_{y+z}(-\frac{1}{2})]\}\\
  &-\frac{\sqrt{2}}{6}\{27[P_{x+y}(\frac{3}{2})-P_{x+y}(-\frac{3}{2})]
  +[P_{x+y}(\frac{1}{2})-P_{x+y}(-\frac{1}{2})]\}\\
&+\frac{\sqrt{3}}{4}\{27[P_{x+y+z}(\frac{3}{2})-P_{x+y+z}(-\frac{3}{2})]
+[P_{x+y+z}(\frac{1}{2})-P_{x+y+z}(-\frac{1}{2})]\}.
\end{aligned}
\end{equation}
\begin{equation}\label{coefficient_13}
\begin{aligned}
 S_{LTT}^{xx} &=
 \frac{\sqrt{2}}{12}\{27[P_{x+z}(\frac{3}{2})-P_{x+z}(-\frac{3}{2})]
 +[P_{x+z}(\frac{1}{2})-P_{x+z}(-\frac{1}{2})]\}\\
 &-\frac{\sqrt{2}}{12}\{27[P_{x-z}(\frac{3}{2})-P_{x-z}(-\frac{3}{2})]
 +[P_{x-z}(\frac{1}{2})-P_{x-z}(-\frac{1}{2})]\}\\
  &-\frac{\sqrt{2}}{12}\{27[P_{y+z}(\frac{3}{2})-P_{y+z}(-\frac{3}{2})]
  +[P_{y+z}(\frac{1}{2})-P_{y+z}(-\frac{1}{2})]\}\\
  &+\frac{\sqrt{2}}{12}\{27[P_{y-z}(\frac{3}{2})-P_{y-z}(-\frac{3}{2})]
  +[P_{y-z}(\frac{1}{2})-P_{y-z}(-\frac{1}{2})]\},
\end{aligned}
\end{equation}
\begin{equation}\label{coefficient_14}
\begin{aligned}
    S_{TTT}^{xxx} &=
   \frac{1}{4}\{27[P_{x}(\frac{3}{2})-P_{x}(-\frac{3}{2})]
   +[P_{x}(\frac{1}{2})-P_{x}(-\frac{1}{2})]\}\\
   &-\frac{\sqrt{2}}{8}\{27[P_{x+y}(\frac{3}{2})-P_{x+y}(-\frac{3}{2})]
   +[P_{x+y}(\frac{1}{2})-P_{x+y}(-\frac{1}{2})]\}\\
   &-\frac{\sqrt{2}}{8}\{27[P_{x-y}(\frac{3}{2})-P_{x-y}(-\frac{3}{2})]
   +[P_{x-y}(\frac{1}{2})-P_{x-y}(-\frac{1}{2})]\},
\end{aligned}
\end{equation}
\begin{equation}\label{coefficient_15}
\begin{aligned}
   S_{TTT}^{yxx} &=
   -\frac{1}{4}\{27[P_{y}(\frac{3}{2})-P_{y}(-\frac{3}{2})]
   +(P_{y}(\frac{1}{2})-P_{y}(-\frac{1}{2}))]\\
   &+\frac{\sqrt{2}}{8}\{27[P_{y+x}(\frac{3}{2})-P_{y+x}(-\frac{3}{2})]
   +[P_{y+x}(\frac{1}{2})-P_{y+x}(-\frac{1}{2})]\}\\
   &+\frac{\sqrt{2}}{8}\{27[P_{y-x}(\frac{3}{2})-P_{y-x}(-\frac{3}{2})]
   +[P_{y-x}(\frac{1}{2})-P_{y-x}(-\frac{1}{2})]\}.
\end{aligned}
\end{equation}

\section{Spin-3/2 basis matrices}
\label{B:Matrices}
In order to describe the density matrix of a spin-3/2 particle, we employ a set of $\Sigma_{\mu}$ matrices, where $\mu$ takes values from 0 to 15. This set consists of 16 4$\times$4 matrices. Here, we provide the explicit expressions for the $\Sigma_{\mu}$ matrices:
\begin{center}
\begin{equation}
\begin{aligned}
  &\Sigma_{0}=\frac{1}{4}\left(
                   \begin{array}{cccc}
                     1 & 0 & 0 & 0 \\
                     0 & 1 & 0 & 0 \\
                     0 & 0 & 1 & 0 \\
                     0 & 0 & 0 & 1 \\
                   \end{array}
                 \right),
   \Sigma_{1}=\frac{1}{10}\left(
                   \begin{array}{cccc}
                     3 & 0 & 0 & 0 \\
                     0 & 1 & 0 & 0 \\
                     0 & 0 & -1 & 0 \\
                     0 & 0 & 0 & -3 \\
                   \end{array}
                 \right),
 \quad \Sigma_{2}=\frac{1}{10}\left(
        \begin{array}{cccc}
          0 & \sqrt{3} & 0 & 0 \\
          \sqrt{3} & 0 & 2 & 0 \\
          0 & 2 & 0 & \sqrt{3} \\
          0 & 0 & \sqrt{3} & 0 \\
        \end{array}
      \right),\\
  & \Sigma_{3}=\frac{i}{10}\left(
        \begin{array}{cccc}
          0 & -\sqrt{3} & 0 & 0 \\
          \sqrt{3} & 0 & -2 & 0 \\
          0 & 2 & 0 & -\sqrt{3} \\
          0 & 0 & \sqrt{3} & 0 \\
        \end{array}
      \right),
      \quad \Sigma_{4}=\frac{1}{4}\left(
                   \begin{array}{cccc}
                     1 & 0 & 0 & 0 \\
                     0 & -1 & 0 & 0 \\
                     0 & 0 & -1 & 0 \\
                     0 & 0 & 0 & 1 \\
                   \end{array}
                 \right),
  \quad \Sigma_{5}=\frac{\sqrt{3}}{12}\left(
                   \begin{array}{cccc}
                     0 & 1 & 0 & 0 \\
                     1 & 0 & 0 & 0 \\
                     0 & 0 & 0 & -1 \\
                     0 & 0 & -1 & 0 \\
                   \end{array}
                 \right),\\
&  \Sigma_{6}=\frac{i\sqrt{3}}{12}\left(
                   \begin{array}{cccc}
                     0 & -1 & 0 & 0 \\
                     1 & 0 & 0 & 0 \\
                     0 & 0 & 0 & 1 \\
                     0 & 0 & -1 & 0 \\
        \end{array}
      \right),
\quad \Sigma_{7}=\frac{\sqrt{3}}{12}\left(
                   \begin{array}{cccc}
                     0 & 0 & 1 & 0 \\
                     0 & 0 & 0 & 1 \\
                     1 & 0 & 0 & 0 \\
                     0 & 1 & 0 & 0 \\
                   \end{array}
                 \right),
  \quad  \Sigma_{8}=\frac{i\sqrt{3}}{12}\left(
                   \begin{array}{cccc}
                     0 & 0 & -1 & 0 \\
                     0 & 0 & 0 & -1 \\
                     1 & 0 & 0 & 0 \\
                     0 & 1 & 0 & 0 \\
        \end{array}
      \right),\\
& \Sigma_{9}=\frac{1}{6}\left(
        \begin{array}{cccc}
                     1 & 0 & 0 & 0 \\
                     0 & -3 & 0 & 0 \\
                    0 & 0 & 3 & 0 \\
                     0 & 0 & 0 & -1 \\
          \end{array}
                 \right),
 \quad \Sigma_{10}=\frac{\sqrt{3}}{6}\left(
        \begin{array}{cccc}
                     0 & 1 & 0 & 0 \\
                     1 & 0 & -\sqrt{3} & 0 \\
                    0 & -\sqrt{3} & 0 & 1 \\
                     0 & 0 & 1 & 0 \\
        \end{array}
      \right),
\quad \Sigma_{11}=\frac{i\sqrt{3}}{6}\left(
        \begin{array}{cccc}
                     0 & -1 & 0 & 0 \\
                     1 & 0 & \sqrt{3} & 0 \\
                    0 & -\sqrt{3} & 0 & -1 \\
                     0 & 0 & 1 & 0 \\
                   \end{array}
                 \right)\\
 & \Sigma_{12}=\frac{\sqrt{3}}{12}\left(
                   \begin{array}{cccc}
                     0 & 0 & 1 & 0 \\
                     0 & 0 & 0 & -1 \\
                     1 & 0 & 0 & 0 \\
                     0 & -1 & 0 & 0 \\
        \end{array}
                 \right),
 \quad \Sigma_{13}=\frac{i\sqrt{3}}{12}\left(
                   \begin{array}{cccc}
                     0 & 0 & -1 & 0 \\
                     0 & 0 & 0 & 1 \\
                     1 & 0 & 0 & 0 \\
                     0 & -1 & 0 & 0 \\
        \end{array}
      \right),
 \quad  \Sigma_{14}=\frac{1}{6}\left(
        \begin{array}{cccc}
                     0 & 0 & 0 & 1 \\
                     0 & 0 & 0 & 0 \\
                    0 & 0 & 0 & 0 \\
                     1 & 0 & 0 & 0 \\
        \end{array}
                 \right)\\
 &\Sigma_{15}= \frac{i}{6}\left(
        \begin{array}{cccc}
                     0 & 0 & 0 & -1 \\
                     0 & 0 & 0 & 0 \\
                    0 & 0 & 0 & 0 \\
                     1 & 0 & 0 & 0 \\
          \end{array}
                 \right).\\
\end{aligned}
\end{equation}
\end{center}

\section{Polarization vectors for spin-1/2, spin-1 and spin-3/2 particles and
antiparticles}\label{C:Polarization vectors}

In the following section, we investigate the polarization state of a particle ($u$) or an antiparticle ($v$) which is characterized by its polarization vector. We consider a particle or antiparticle with a mass $M$ and a four-momentum
\begin{eqnarray}
p^{\mu} & = & \left(E,\left|\vec{p}\right|\sin\theta\cos\phi,\left|\vec{p}\right|\sin\theta\sin\phi,\left|\vec{p}\right|\cos\theta\right).
\end{eqnarray}
For spin-1/2, spin-1, and spin-3/2 particles and antiparticles, we adopt the convention of Pauli-Dirac representation.

The spin-1/2 field is derived by solving the Dirac equation, as detailed in Ref.~\cite{Dirac:1928hu}.
The polarization vectors for spin-1/2 particles with helicity $\lambda=\pm$ are given by:
\begin{align}
u\left(\vec{p},\pm\right) & =  \left(\begin{array}{c}
\sqrt{E+m}\chi_{\pm}\\
\pm\sqrt{E-m}\chi_{\pm}
\end{array}\right),\\
v\left(\vec{p},\pm\right) & =  \left(\begin{array}{c}
\sqrt{E-m}\chi_{\mp}\\
\mp\sqrt{E+m}\chi_{\mp}
\end{array}\right),
\end{align}
where $\chi_{\pm}$ represent the two-component spinors:
\begin{align}
\chi_{+} & =  \left(\begin{array}{c}
\cos\frac{\theta}{2}\\
\sin\frac{\theta}{2}e^{i\phi}
\end{array}\right),
\chi_{-}  =  \left(\begin{array}{c}
-\sin\frac{\theta}{2}\\
\cos\frac{\theta}{2}e^{i\phi}
\end{array}\right).
\end{align}

The spin-1 field is derived by solving the Proca equation, as outlined in Ref.~\cite{Proca:1936fbw}. The polarization vectors for spin-1 particles with helicity $\lambda=\pm,0$ are given by:
\begin{eqnarray}
\varepsilon^{\mu}\left(\vec{p},\pm\right) & = & \frac{1}{\sqrt{2}}\left(\begin{array}{c}
0\\
\mp\cos\theta\cos\phi+i\sin\phi\\
\mp\cos\theta\sin\phi-i\cos\phi\\
\pm\sin\theta
\end{array}\right),\\
\varepsilon^{\mu}\left(\vec{p},0\right) & = & \frac{1}{m}\left(\begin{array}{c}
\left|\vec{p}\right|\\
E\sin\theta\cos\phi\\
E\sin\theta\sin\phi\\
E\cos\theta
\end{array}\right).
\end{eqnarray}

The spin-3/2 field is derived by solving the Rarita-Schwinger equation, as outlined in Ref.~\cite{Rarita:1941mf}. The polarization vectors for spin-3/2 particles with helicity $\lambda=\pm\frac{3}{2},\pm\frac{1}{2}$ are given by:
\begin{align}
u^{\mu}\left(p_{z},\pm\frac{3}{2}\right) & =  \varepsilon_{\pm}^{\mu}\left(p_{z}\right)u_{\pm}\left(p_{z}\right),\\
u^{\mu}\left(p_{z},\pm\frac{1}{2}\right) & =  a\varepsilon_{\pm}^{\mu}\left(p_{z}\right)u_{\mp}\left(p_{z}\right)+b\varepsilon_{0}^{\mu}\left(p_{z}\right)u_{\pm}\left(p_{z}\right),\\
v^{\mu}\left(p_{z},\pm\frac{3}{2}\right) & =  \varepsilon_{\pm}^{\mu*}\left(p_{z}\right)v_{\pm}\left(p_{z}\right),\\
v^{\mu}\left(p_{z},\pm\frac{1}{2}\right) & =  a\varepsilon_{\pm}^{\mu*}\left(p_{z}\right)v_{\mp}\left(p_{z}\right)+b\varepsilon_{0}^{\mu*}\left(p_{z}\right)v_{\pm}\left(p_{z}\right),
\end{align}
where $a$ and $b$ are normalization coefficients that satisfy $a^{2}+b^{2}=1$. In the context of the SU(6) quark model, the values $a=\sqrt{1/3}$ and $b=\sqrt{2/3}$ are determined to remove the spin-1/2 fermions.

\section{scattering amplitudes for $e^{+}e^{-}\rightarrow\Omega^{-}\bar{\Omega}^{+}$}\label{D:scattering amplitudes}

When investigating the longitudinal polarization dependence of the reaction cross-section, we compute the helicity scattering amplitude. Considering the principles of parity conservation and charge conjugation invariance, we derive the following independent terms:

\begin{align}
\left|\mathcal{M}_{+,-,\pm\frac{3}{2},\pm\frac{3}{2}}\right|^{2} & =\left|\mathcal{M}_{-,+,\pm\frac{3}{2},\pm\frac{3}{2}}\right|^{2}=e^{4}\frac{m^{2}}{E^{2}}\sin^{2}\theta,\\
\left|\mathcal{M}_{+,-,\pm\frac{1}{2},\pm\frac{1}{2}}\right|^{2} & =\left|\mathcal{M}_{-,+,\pm\frac{1}{2},\pm\frac{1}{2}}\right|^{2}=e^{4}\left[\frac{2b^{2}E^{2}-\left(a^{2}+b^{2}\right)m^{2}}{mE}\right]^{2}\sin^{2}\theta,\\
\left|\mathcal{M}_{+,-,\pm\frac{1}{2},\mp\frac{1}{2}}\right|^{2} & =\left|\mathcal{M}_{-,+,\mp\frac{1}{2},\pm\frac{1}{2}}\right|^{2}=e^{4}b^{4}\left(\frac{2E^{2}-m^{2}}{m^{2}}\right)^{2}\left(1\mp\cos\theta\right)^{2},\\
\left|\mathcal{M}_{+,-,\pm\frac{3}{2},\pm\frac{1}{2}}\right|^{2} & =\left|\mathcal{M}_{-,+,\mp\frac{3}{2},\mp\frac{1}{2}}\right|^{2}=\left|\mathcal{M}_{-,+,\pm\frac{1}{2},\pm\frac{3}{2}}\right|^{2}=\left|\mathcal{M}_{+,-,\mp\frac{1}{2},\mp\frac{3}{2}}\right|^{2}=e^{4}a^{2}\left(1\mp\cos\theta\right)^{2}.
\end{align}

When investigating the cross-section's dependence on the transverse polarization of the $\Omega^{-}$ particle, we decompose its polarization in the transverse direction. For the other particles, we sum over their polarizations, allowing for longitudinal polarization decomposition. By utilizing Eq.~\eqref{direction_Axis} and considering parity conservation, we derive the following independent terms:
\begin{align}
\left|\mathcal{M}_{+,-,\pm\frac{3}{2}\uparrow_{x},+\frac{3}{2}}\right|^{2}  =&\left|\mathcal{M}_{-,+,\mp\frac{3}{2}\uparrow_{x},-\frac{3}{2}}\right|^{2}=\frac{e^{4}}{8E^{2}}\left[\sqrt{3}aE\left(1+\cos\theta\right)\pm m\sin\theta\right]^{2},\\
\left|\mathcal{M}_{-,+,\pm\frac{3}{2}\uparrow_{x},+\frac{3}{2}}\right|^{2}  =&\left|\mathcal{M}_{+,-,\mp\frac{3}{2}\uparrow_{x},-\frac{3}{2}}\right|^{2}=\frac{e^{4}}{8E^{2}}\left[\sqrt{3}aE\left(1-\cos\theta\right)\mp m\sin\theta\right]^{2},\\
\left|\mathcal{M}_{+,-,\pm\frac{3}{2}\uparrow_{x},+\frac{1}{2}}\right|^{2}  =&\left|\mathcal{M}_{-,+,\mp\frac{3}{2}\uparrow_{x},-\frac{1}{2}}\right|^{2}=\frac{e^{4}}{8E^{2}m^{4}}\left\{ am^{2}\left[E\left(1-\cos\theta\right)\pm\sqrt{3}am\sin\theta\right]\right.\nonumber\\
 & \qquad\left.-\sqrt{3}b^{2}\left(2E^{2}-m^{2}\right)\left[E\left(1+\cos\theta\right)\pm m\sin\theta\right]\right\} ^{2},\\
\left|\mathcal{M}_{-,+,\pm\frac{3}{2}\uparrow_{x},+\frac{1}{2}}\right|^{2}  =&\left|\mathcal{M}_{+,-,\mp\frac{3}{2}\uparrow_{x},-\frac{1}{2}}\right|^{2}=\frac{e^{4}}{8E^{2}m^{4}}\left\{ am^{2}\left[E\left(1+\cos\theta\right)\mp\sqrt{3}am\sin\theta\right]\right.\nonumber\\
 & \qquad\left.-\sqrt{3}b^{2}\left(2E^{2}-m^{2}\right)\left[E\left(1-\cos\theta\right)\mp m\sin\theta\right]\right\} ^{2},\\
\left|\mathcal{M}_{+,-,\pm\frac{1}{2}\uparrow_{x},+\frac{3}{2}}\right|^{2}  =&\left|\mathcal{M}_{-,+,\mp\frac{1}{2}\uparrow_{x},-\frac{3}{2}}\right|^{2}=\frac{e^{4}}{8E^{2}}\left[aE\left(1+\cos\theta\right)\pm\sqrt{3}m\sin\theta\right]^{2},\\
\left|\mathcal{M}_{-,+,\pm\frac{1}{2}\uparrow_{x},+\frac{3}{2}}\right|^{2}  =&\left|\mathcal{M}_{+,-,\mp\frac{1}{2}\uparrow_{x},-\frac{3}{2}}\right|^{2}=\frac{e^{4}}{8E^{2}}\left[aE\left(1-\cos\theta\right)\mp\sqrt{3}m\sin\theta\right]^{2},\\
\left|\mathcal{M}_{+,-,\pm\frac{1}{2}\uparrow_{x},+\frac{1}{2}}\right|^{2}  =&\left|\mathcal{M}_{-,+,\mp\frac{1}{2}\uparrow_{x},-\frac{1}{2}}\right|^{2}=\frac{e^{4}}{8E^{2}m^{4}}\left\{ am^{2}\left[\sqrt{3}E\left(1-\cos\theta\right)\pm am\sin\theta\right]\right.\nonumber\\
 & \qquad\left.+b^{2}\left(2E^{2}-m^{2}\right)\left[E\left(1+\cos\theta\right)\mp m\sin\theta\right]\right\} ^{2},\\
\left|\mathcal{M}_{-,+,\pm\frac{1}{2}\uparrow_{x},+\frac{1}{2}}\right|^{2}  =&\left|\mathcal{M}_{+,-,\mp\frac{1}{2}\uparrow_{x},-\frac{1}{2}}\right|^{2}=\frac{e^{4}}{8E^{2}m^{4}}\left\{ am^{2}\left[\sqrt{3}E\left(1+\cos\theta\right)\mp am\sin\theta\right]\right.\nonumber\\
 & \qquad\left.+b^{2}\left(2E^{2}-m^{2}\right)\left[E\left(1-\cos\theta\right)\pm m\sin\theta\right]\right\} ^{2},\\
 \left|\mathcal{M}_{+,-,\pm\frac{3}{2}\uparrow_{y},+\frac{3}{2}}\right|^{2}  =&\left|\mathcal{M}_{-,+,\mp\frac{3}{2}\uparrow_{y},-\frac{3}{2}}\right|^{2}=\frac{e^{4}}{8E^{2}}\left[3a^{2}E^{2}\left(1+\cos\theta\right)^{2}+m^{2}\sin^{2}\theta\right]\\
\left|\mathcal{M}_{-,+,\pm\frac{3}{2}\uparrow_{y},+\frac{3}{2}}\right|^{2}  =&\left|\mathcal{M}_{+,-,,\mp\frac{3}{2}\uparrow_{y},-\frac{3}{2}}\right|^{2}=\frac{e^{4}}{8E^{2}}\left[3a^{2}E^{2}\left(1-\cos\theta\right)^{2}+m^{2}\sin^{2}\theta\right]\\
\left|\mathcal{M}_{+,-,\pm\frac{3}{2}\uparrow_{y},+\frac{1}{2}}\right|^{2}  =&\left|\mathcal{M}_{-,+,\mp\frac{3}{2}\uparrow_{y},-\frac{1}{2}}\right|^{2}=\frac{e^{4}}{8E^{2}m^{4}}\left\{ 3m^{2}\left[m^{2}\left(a^{2}+b^{2}\right)-2b^{2}E^{2}\right]^{2}\sin^{2}\theta\right.\nonumber\\
 & \qquad\left.+\left[am^{2}E\left(1-\cos\theta\right)+\sqrt{3}b^{2}E\left(2E^{2}-m^{2}\right)\left(1+\cos\theta\right)\right]^{2}\right\} \\
\left|\mathcal{M}_{-,+,\pm\frac{3}{2}\uparrow_{y},+\frac{1}{2}}\right|^{2}  =&\left|\mathcal{M}_{+,-,\mp\frac{3}{2}\uparrow_{y},-\frac{1}{2}}\right|^{2}=\frac{e^{4}}{8E^{2}m^{4}}\left\{ 3m^{2}\left[m^{2}\left(a^{2}+b^{2}\right)-2b^{2}E^{2}\right]^{2}\sin^{2}\theta\right.\nonumber\\
 & \qquad\left.+\left[am^{2}E\left(1+\cos\theta\right)+\sqrt{3}b^{2}E\left(2E^{2}-m^{2}\right)\left(1-\cos\theta\right)\right]^{2}\right\} \\
\left|\mathcal{M}_{+,-,\pm\frac{1}{2}\uparrow_{y},+\frac{3}{2}}\right|^{2}  =&\left|\mathcal{M}_{-,+,\mp\frac{1}{2}\uparrow_{y},-\frac{3}{2}}\right|^{2}=\frac{e^{4}}{8E^{2}}\left[a^{2}E^{2}\left(1+\cos\theta\right)^{2}+3m^{2}\sin^{2}\theta\right],\\
\left|\mathcal{M}_{-,+,\pm\frac{1}{2}\uparrow_{y},+\frac{3}{2}}\right|^{2}  =&\left|\mathcal{M}_{+,-,\mp\frac{1}{2}\uparrow_{y},-\frac{3}{2}}\right|^{2}=\frac{e^{4}}{8E^{2}}\left[a^{2}E^{2}\left(1-\cos\theta\right)^{2}+3m^{2}\sin^{2}\theta\right],\\
\left|\mathcal{M}_{+,-,\pm\frac{1}{2}\uparrow_{y},+\frac{1}{2}}\right|^{2}  =&\left|\mathcal{M}_{-,+,\mp\frac{1}{2}\uparrow_{y},-\frac{1}{2}}\right|^{2}=\frac{e^{4}}{8E^{2}m^{4}}\left\{ m^{2}\left[m^{2}\left(a^{2}+b^{2}\right)-2b^{2}E^{2}\right]^{2}\sin^{2}\theta\right.\nonumber\\
 & \qquad\left.+\left[\sqrt{3}am^{2}E\left(1-\cos\theta\right)-b^{2}E\left(2E^{2}-m^{2}\right)\left(1+\cos\theta\right)\right]^{2}\right\} ,
\end{align}
\begin{align}
\left|\mathcal{M}_{-,+,\pm\frac{1}{2}\uparrow_{y},+\frac{1}{2}}\right|^{2}  =&\left|\mathcal{M}_{+,-,\mp\frac{1}{2}\uparrow_{y},-\frac{1}{2}}\right|^{2}=\frac{e^{4}}{8E^{2}m^{4}}\left\{ m^{2}\left[m^{2}\left(a^{2}+b^{2}\right)-2b^{2}E^{2}\right]^{2}\sin^{2}\theta\right.\nonumber\\
 & \qquad\left.+\left[\sqrt{3}am^{2}E\left(1+\cos\theta\right)-b^{2}E\left(2E^{2}-m^{2}\right)\left(1-\cos\theta\right)\right]^{2}\right\}.
 \end{align}

\end{CJK*}
\end{document}